\documentclass[amsmath,amssymb,prb,twocolumn]{revtex4-1}

\usepackage[utf8]{inputenc}
\usepackage{graphicx}
\usepackage{dcolumn}%
\usepackage{bm}
\usepackage{appendix}
\usepackage{braket}
\usepackage{color}
\usepackage{hyperref}
\usepackage{comment}

\graphicspath{{./Figures/}}

\DeclareMathOperator{\Tr}{Tr}
\begin{document}

\title{Relativistic magnetic interactions 
from non-orthogonal basis sets}

\author{Gabriel Martínez-Carracedo$^{1,2}$}
\author{László Oroszlány$^{3,4}$}
\author{Amador García-Fuente$^{1,2}$}
\author{Bendegúz Nyári$^{5,6}$}
\author{László Udvardi$^{6}$}
\author{László Szunyogh$^{5,6}$}
\author{Jaime Ferrer$^{1,2}$}

\affiliation{
$^1$Departamento de Física,  Universidad de Oviedo,  33007 Oviedo, Spain\\
$^2$Centro de Investigación en Nanomateriales y Nanotecnología, Universidad de Oviedo-CSIC, 33940, El Entrego, Spain\\
$^3$Department of Physics of Complex Systems, ELTE E\"otv\"os Lor\'and University, H-1117, Budapest,
Hungary\\
$^4$Wigner Research Centre for Physics, H-1525, Budapest, Hungary\\
$^5$HUN-REN-BME Condensed Matter Research Group, Budapest University of Technology and Economics, 
M\H{u}egyetem rkp. 3., H-1111 Budapest, Hungary \\
$^6$Department of Theoretical Physics, Institute of Physics, Budapest University of Technology and Economics, 
M\H{u}egyetem rkp. 3., H-1111 Budapest, Hungary
}

\begin{abstract}
We propose a method to determine the magnetic exchange interaction and on-site 
anisotropy tensors  
of extended Heisenberg spin models from density functional theory 
including relativistic effects. 
The method is based on the 
Liechtenstein-Katsnelson-Antropov-Gubanov torque formalism, whereby energy variations upon infinitesimal 
rotations are 
performed. We assume that the Kohn-Sham Hamiltonian is expanded in a non-orthogonal basis set of pseudo-atomic
orbitals. We define local operators that are both hermitian and satisfy relevant sum rules. 
We demonstrate that in the presence of spin-orbit coupling a correct mapping from the density functional total energy to a spin model 
that relies on the rotation of the exchange field part of the Hamiltonian can not be accounted for by transforming the full Hamiltonian. We derive a set of sum rules that pose stringent validity tests on any 
specific calculation. We showcase the flexibility and accuracy of the method by computing the exchange and anisotropy 
tensors of both well-studied magnetic nanostructures and of recently synthesized  two-dimensional magnets. Specifically, we benchmark 
our approach against the established Korringa-Kohn-Rostoker Green's function method and show that they agree well. Finally, we demonstrate how the application 
of biaxial strain on the two-dimensional magnet T-CrTe$_2$ can trigger 
a magnetic phase transition.   
\end{abstract}

\maketitle


\section{Introduction}
The discovery of magnetism in the van der Waals materials CrI$_3$ and Cr$_2$Ge$_2$Te$_6$ \cite{Huang17,Gong17} has 
unleashed an intense theoretical and experimental activity on 2-dimensional and layer-dependent magnetism, as well 
as raised expectations on the development of advanced magnetic, spintronic, magneto-optical, magnetocalorimetric and 
quantum technologies\cite{Wang22}. Efforts to raise the critical temperature beyond room temperature, as well as to
achieve a large perpendicular anisotropy are currently being undertaken, with promising candidates such as 
Fe$_3$GaTe$_2$ \cite{Zhang2022}. Almost all van der Waals magnets involve a magnetic transition metal atom together
with a halogen such as the trihalides CrX$_3$ (X=Br, Cl and I) \cite{C9CP01837A,PhysRevLett.128.177202}, or a 
chalcogen such as MPX$_3$ (M=Mn, Fe, Ni; X=S, Se) \cite{LEFLEM1982455,PhysRevB.46.5425} or 
MoTe$_2$ \cite{Tiwari2021}. Some magnets also feature a more complex stoichiometry and crystal structure like 
Fe$_3$GeTe$_2$ \cite{Zhang2022,Roemer2020}.

The magnetic response of most magnetic materials can be derived from the generalized classical Heisenberg 
model
\begin{eqnarray}
\label{eqn:Heisenberg1}
H(\{ {\bf S}_i\}) &=&\,\frac{1}{2}\,\sum_{i\neq j}\,{\bf S}_i\,{\cal J}_{ij}\,{\bf S}_j+\
\sum_i\,{\bf S}_i\,{\cal K}_i\,{\bf S}_i
\end{eqnarray}
where ${\bf S}_i=\hbar\,S_i\,{\bf e}_i$ indicates the angular momentum vector of an atom or localized 
magnetic entity placed at site $i$ in the material,  whose modulus is $\hbar\,S_i$ and whose direction is
given by the unit vector ${\bf e}_i$. ${\cal J}_{ij}$ and ${\cal K}_i$ are $3\times 3$ tensors 
that describe the exchange coupling between sites $i$ and $j$, and the intra-atomic magnetic anisotropy
at site $i$, respectively.  The above Hamiltonian can also be rewritten as
\begin{eqnarray}
\label{eqn:Heisenberg2}
H(\{ {\bf e}_i\})
&=&\,\frac{1}{2}\,\sum_{i\neq j}\,{\bf e}_i\,J_{ij}\,{\bf e}_j+\sum_i\,{\bf e}_i\,K_i\,{\bf e}_i
\end{eqnarray}
by defining the renormalized tensors $J_{ij}=\hbar^2 S_i S_j\,{\cal J}_{ij}$ and $K_i=(\hbar S_i)^2\,{\cal K}_i$,
that have energy units. These tensors satisfy that $J_{ij}=J_{ji}^T$ and $K_i=K_i^T$, where $T$ denotes 
the transpose matrix.  It is therefore pressing to develop adequate theoretical and experimental tools that 
determine the above exchange and anisotropy tensors and use them to gain a deeper understanding and control of 
layered magnetism.

Density Functional Theory (DFT) \cite{HK,KS} is a platform, based on a fully quantum-mechanical approach, 
upon which the above tools may be developed.
The reason behind this assertion is that DFT delivers the exact ground-state energy $E_G$ and 
density $n_G(\bf{r})$ of any material, provided that the exact Exchange-Correlation (XC) functional is 
known\cite{CFMB}. DFT establishes that there exists a functional $F[n]$ of the electron density 
$n({\bf r})=\frac{1}{2}\,(n_0({\bf r}) \,\tau_0+\bf{m}(\bf{r})\cdot {\boldsymbol \tau})$ which, when evaluated at the exact 
ground state density $n_G({\bf r})$, delivers the ground-state energy $E_G=F[n_G]$.  Here, $n({\bf r})$ is a $
2\times 2$ matrix in spin space  where $n_0({\bf r})$ and ${\bf m}(\bf{r})$ are the charge and magnetization 
densities, $\tau_0$ is the $2\times 2$ identity matrix and ${\boldsymbol \tau}$ denotes the vector of 
Pauli matrices acting on the spin degree of freedom. Therefore, the $F-$functional can also be written as $F[n_0;{\bf m}]$ and the 
ground state energy is $E_G=F[n_{0,G};{\bf m}_G]$. It is hence natural whenever simulating a 
magnetic system, to identify the localized spin momentum vector ${\bf S}_i=S_i\,{\bf e}_i$ in 
Eqs.  (\ref{eqn:Heisenberg1}) and (\ref{eqn:Heisenberg2}) with a suitable average of the ground-state 
magnetization ${\bf m}_G({\bf r})$ over a given spatial region {\it i} so that the ground-state energy becomes 
effectively a function of ${\bf S}_i$.  Then,  if $S_i$ is found to be constant, $F[n_{0,G};{\bf e}_i]$ can 
be expanded in powers of ${\bf e}_i$, thereby providing a mapping to the global energy minimum of 
the classical Hamiltonian (\ref{eqn:Heisenberg2}), from which 
the magnetic parameters can be extracted. 
This mapping can be drawn not only between the ground states of the classical Hamiltonian and the 
$F-$functional, but also between other pairs of local minima. 
In this work we will call this parallelism the DFT-to-Spin Model (DFT2S) mapping.

A meaningful mapping can only be performed for electron states that are sufficiently localized within 
given atoms and whose charge and longitudinal spin degrees of freedom are frozen so that their total
spin is sufficiently well-defined. A simple example relates to the Hilbert space of a single site or orbital, that 
is composed of four possible states ($\ket{0},\,\ket{\uparrow},\, \ket{\downarrow},\,\ket{\uparrow \downarrow}$). 
A faithful DFT2S mapping can only be achieved if this space can be reduced to the Hilbert space of a spin-$1/2$, that has only the states $\ket{\uparrow}$ and $\ket{\downarrow}$.
The reduction is physically meaningful if the corresponding charge and longitudinal spin correlation functions
have a gap in their excitation spectrum as is illustrated in Ref. [\onlinecite{JF95}] for the 
one-dimensional Hubbard model. We shall argue later in this article that the above Hilbert 
space mismatch is the source of serious numerical issues  
that are only solved when implementing a Hilbert-space reduction from the full variational basis
to the set of localized magnetic orbitals.

Several strategies to carry out a DFT2S mapping have been used in the past. One of the methods consists of 
determining $E_G$ for ferromagnetic and antiferromagnetic minima of the functional, or for orientations 
of the moments along other different directions\cite{LN,FF07}. Then, by computing 
total energy differences, averaged estimates of the magnetic tensors can be retrieved.  The method is however very 
limited because one needs to determine the total energy of many different magnetic configurations in 
order to evaluate and disentangle all components of the exchange matrices, and to differentiate among 
the many neighbors of a given site. Furthermore, for molecules or isolated nanostructures 
only magnetic configurations belonging to the same 
angular momentum multiplet should be compared.

We shall follow instead the {\it torque} or Liechtenstein-Katsnelson-Antropov-Gubanov (LKAG) 
method \cite{LKAG,Katsnelson_Jij_TBcorr,Katsnelson_Jij_Wannier,KatsnelsonPRB2010,arxivpaper}, 
that provides accurate estimates of the full magnetic tensors with a fraction of the cost required by the total 
energy differences method. The LKAG method relies on the so-called magnetic force theorem \cite{MFT}, which
establishes that upon application of an infinitesimal perturbation to the ground state of the 
many-electron system, its total energy variation is equal to the variation of the ground-state energy of the 
Kohn-Sham (KS) Hamiltonian. The LKAG approach then consists of 
performing suitable infinitesimal rotations of the angular momenta of both the classical and the KS Hamiltonians, 
on calculating the corresponding energy differences by use of second-order perturbation theory, and on 
equating those second-order energy variations.

We mention here the LKAG implementation performed in the 
Korringa--Kohn--Rostoker Green's function (KKR-GF) \cite{KKRBOOK} and the tight-binding linear 
muffin-tin orbital (TB-LMTO) methods \cite{TBLMTO-Andersen,TBLMTO-Turek}, that have been particularly 
successful for calculating magnetic exchange interactions of a number of bulk materials, surfaces, 
interfaces, films, superlattices, or finite metallic clusters in the past 
\cite{Ebert2011,Pajda2001,Pajda2000}. Furthermore, the calculation of tensorial exchange interactions, 
including two-ion magnetic anisotropy and DM vectors, has become available by extending the LKAG formula 
to the relativistic case \cite{Udvardi_relativistic_Jij,ebertPRB09}. This extension opened the door to 
the analysis, design and tuning of complex magnetic states like domain walls \cite{Vida-domainwalls}, 
spin spirals \cite{Rozsa-FeWTa,Simon-cappedCoPt} and magnetic skyrmions 
\cite{Polesya-FeTaS-skyrmion,Rozsa-skyrmion1,Rozsa-skyrmion2,Hsu-HFePt-skyrmions} in ultrathin films. 
However, the  KKR-GF or TB-LMTO methods are not designed to  describe open systems or generically 
systems lacking high degrees of symmetry such as nano-scale structures. 

A torque 
DFT2S mapping is facilitated if the starting electron Hamiltonian is written in 
the tight-binding language,  where the electron degrees of freedom are already expressed in terms 
of localized orbitals. A plane-wave basis might be chosen as a starting point, and the ensuing 
eigenstates be {\it wannierized} afterwards yielding a tight-binding Hamiltonian. 
However, it turns out that those Wannier states may not be centered close enough to the physical atoms
\cite{Maximally_Localized_Wannier_RMP,TB2J}, raising difficulties on the DFT2S mapping if this is the case.

This difficulty is overcome if the physical states are expressed in terms of a Pseudo-Atomic-Orbitals (PAO) basis set.
PAO-based implementations of DFT, such as the SIESTA method\cite{siestapaper}, have proven to be a flexible,
fast and cost-effective approach for studying generic nanostructures.
We have provided recently a generalization of the non-relativistic LKAG DFT2S mapping to PAO
bases\cite{Oroszlany2019}, therefore gaining access to the isotropic Heisenberg exchange constants $J^H_{ij}$, 
that are proportional to the trace of
the exchange tensor  in Eq. (\ref{eqn:Heisenberg2}). We have also demonstrated the usefulness of the new 
approach for several graphene based materials\cite{Carracedo2023}. 
The PAO approach has however the inconvenience that the concept of a localized operator is a non-trivial 
issue \cite{Juanjo14} that has not yet been resolved. 

We present in this manuscript an extension of our previous work to the fully relativistic case, that enables
us to determine all the matrix elements of the $J_{ij}$ and $K_i$ tensors in Eq. (\ref{eqn:Heisenberg2}).
We establish several sum rules related to energy variations on the one hand, and exchange/anisotropy tensors
on the other, that apply for specific symmetry operations.
In addition, we propose a new definition of a local operator, that is suitable for both orthogonal 
and non-orthogonal basis sets, and fulfills those sum rules. We demonstrate that in presence of 
spin-orbit coupling a DFT2S mapping can not be achieved by performing spin or total angular momentum rotations on the 
full Hamiltonian. 
We also find a quantum analog of the Steiner theorem,
that relates off-site to on-site matrix elements of the orbital angular momentum. 
We demonstrate the accuracy and flexibility of the
approach by its successful description of the exchange and anisotropy parameters of two small atomic clusters, 
of two heterostructures, and of the van der Waals magnet T-CrTe$_2$.

The layout of this article is as follows. Section \ref{section:PAObasis} describes how to deploy the quantum-mechanical 
algebra needed to develop the DFT2S mapping using an non-orthogonal basis set.
Section \ref{section:Perturbativeclassical} describes how to perform local infinitesimal rotations of the angular 
moments in Eq. (\ref{eqn:Heisenberg2}), and compute energy variations up to second order in perturbation theory.
Section \ref{section:KS} describes general principles on how to perform analogous infinitesimal angular 
moment variations in the DFT system, and describes our specific implementation of the DFT2S mapping.
Section \ref{section:Tests} describes the tests that we have performed to validate our method. We show
that our approach satisfies several stringent consistency checks and sum rules. We also discuss the
benchmarks that we have performed against the 
well-established KKR-GF method for a number of magnetic heterostructures. Section 
\ref{section:Crte2} describes the exchange and anisotropy tensors of the van der Waals magnet T-CrTe$_2$, 
and demonstrates that
a magnetic phase transitions can be driven and controlled by biaxial strain. 

An implementation of the LKAG-PAO approach called TB2J \cite{TB2J} has also been proposed recently.
TB2J has been substantiated as a 
post-processing tool that interfaces both to the Wannier90 \cite{wannierf90} and the SIESTA codes. Our approach 
differs from the TB2J implementation in several crucial aspects. We discuss those differences in sections 
\ref{section:Perturbativeclassical} and \ref{section:KS}. We have also tested the impact of those differences
on the quality of the determined exchange and anisotropy tensors in section \ref{section:Tests}.

\section{Quantum mechanics using a non-orthogonal basis set}
\label{section:PAObasis}
A DFT2S mapping between the Heisenberg Hamiltonian in Eq. ({\ref{eqn:Heisenberg2}) and a KS 
Hamiltonian written in the tight-binding language is appealing because the physical
magnitudes in both cases are centered at atoms. A KS tight-binding Hamiltonian can be
obtained by the use of Wannier orthonormal orbitals. However, while these orbitals are
rather localized they might not be centered close enough to atomic sites. In contrast, 
PAO basis functions are strictly centered at atomic sites but are not orthogonal, which leads to 
difficulties when defining local operators as explained in detail by Soriano and 
Palacios\cite{Juanjo14}. In order to provide with a coherent discussion we reproduce here a considerable part of their development.  

We devote this section to review the algebraic structure needed to perform quantum-mechanical algebra
using a non-orthogonal basis set. Large swathes of the section are known\cite{siestapaper,Juanjo14},
while other are non-trivial, and specifically, the definition of a local operator,
and the correct way to perform symmetry operations has not been discussed in the past 
to the best of our knowledge.

\subsection{PAO basis}
Let us suppose that the Hilbert space ${\cal E}_N$ of physical states of a quantum system 
is variationally spanned in the PAO basis set $E= \{\ket{\mu_1},\ket{\mu_2},\dots, \ket{\mu_N}\}$,
such that the scalar product between two PAOs is the overlap integral $O_{12}=\braket{\mu_1|\mu_2}$, and $N$ 
is the basis size. The overlap integrals can be arranged to form the $N\times N$ matrix $O$.

Let ${\cal E}^{*}_N$ be the dual space where we define the dual basis  set
$E\text{*}=\{\,\ket{\mu_1\text{*}},\,\ket{\mu_2\text{*}},\dots, \ket{\mu_N\text{*}}\,\}$ 
such that the orthogonality relation $\braket{\mu|\nu\text{*}}=\braket{\mu\text{*}|\nu}=\delta_{\mu\nu}$ is fulfilled.
Then the scalar product between two dual vectors is $\braket{\mu\text{*}|\nu\text{*}}=\left(O^{-1}\right)_{\mu\nu}$.

We introduce now the following compressed notation. We arrange the PAO and dual PAO bases in row vectors of kets
\begin{equation}
\begin{array}{l}
|\,\epsilon\rangle\rangle=\left(\,\ket{\mu_1},\ket{\mu_2},\dots,\,\ket{\mu_N}\,\right)\\ [5pt]
|\,\epsilon\text{*}\rangle\rangle=\left(\,\ket{\mu_1\text{*}},\ket{\mu_2\text{*}},\dots,\,\ket{\mu_N\text{*}}\,\right)
\end{array} 
\end{equation}
so that the overlap and orthogonality relation can be expressed compactly as
\begin{equation}
\begin{array}{lcl}
\langle\langle\epsilon\,|\,\epsilon\rangle\rangle&=&O\\ [5pt]
\langle\langle\epsilon\text{*}\,|\,\epsilon\text{*}\rangle\rangle&=&O^{-1} \\ [5pt]
\langle\langle\epsilon\text{*}\,|\,\epsilon\rangle\rangle&=&\langle\langle\epsilon\,|\,\epsilon\text{*}\rangle\rangle=I_N
\end{array}
\end{equation}
where $I_N$ is the $N\times N$ identity matrix. 
Notice that the dual and direct bases become the same when the overlap 
matrix tends to the identity matrix. The closure relation can be expressed compactly as
\begin{equation}
\begin{array}{lcl}
\hat{I}&=&|\,\epsilon\rangle\rangle\langle\langle\epsilon\text{*}\,|= \displaystyle \sum_{\mu\in{\cal E}}\ket{\mu}\bra{\mu\text{*}}\\ [15pt]
&=&|\,\epsilon\text{*}\rangle\rangle\langle\langle\epsilon|\,= \displaystyle \sum_{\mu\in{\cal E}}\ket{\mu\text{*}}\bra{\mu}
\end{array}
\end{equation}
where $\hat{I}$ is the identity operator.

Let us assume now that we are interested in a PAO basis subset that we call $A=\{\ket{\mu_1},\ket{\mu_2},\dots, \ket{\mu_A}\}$ and let us call 
$B=\{\ket{\mu_{A+1}},\ket{\mu_{A+2}},\dots, \ket{\mu_N}\}$ the complementary subset such that $E=A \cup B$. We arrange the two subsets
into the row vectors of kets 
\begin{equation}
\begin{array}{ll}
|\,A\rangle\rangle&=\left(\,\ket{\mu_1},\ket{\mu_2},\dots,\ket{\mu_A}\,\right)\\ [5pt]
|\,B\rangle\rangle&=\left(\,\ket{\mu_{A+1}},\ket{\mu_{A+2}},\dots,\ket{\mu_N}\,\right)
\end{array}
\end{equation}
We can define the following non-hermitian pseudo-projection operators 
\begin{equation}
\begin{array}{lcl}
\hat{P}_A=|\,A\rangle\rangle\,\langle\langle A\text{*}\,|&\neq&\hat{P}_A^\dagger=|\,A\text{*}\rangle\rangle\,\langle\langle A\,|\\ [5pt]
\hat{P}_B=|\,B\rangle\rangle\,\langle\langle B\text{*}\,|&\neq&\hat{P}_B^\dagger=|\,B\text{*}\rangle\rangle\,\langle\langle B\,|
\end{array}
\end{equation}
that fulfill the closure relation
\begin{equation}
\hat{I}=\hat{P}_A+\hat{P}_B=\hat{P}_A^\dagger+\hat{P}_B^\dagger
\end{equation}
These operators also satisfy that $\hat{P}_A=\hat{P}_A^2$ but are not true
projectors because they are not hermitian and because $\hat{P}_A^\dagger\,\hat{P}_B$ is not equal to the  null operator.

\subsection{Operators and local operators}
We discuss now how to span linear operators acting on ${\cal E}_N$, ${\cal E}^*_N$ in the PAO basis. 
Let $\hat{C}$ be one such operator. We can express $\hat{C}$ either by computing matrix 
elements in the direct basis $C_{\mu\nu}=\braket{\mu|\hat{C}|\nu}$, or in the dual basis
$C_{\mu\nu}^{*}=\braket{\mu\text{*}|\hat{C}|\nu\text{*}}$:
\begin{equation}
\begin{array}{lcl}
\hat{C}&=& \displaystyle \sum_{\mu,\nu\in E,E^{*}}\ket{\mu\text{*}}C_{\mu\nu}\bra{\nu\text{*}}=|\,\epsilon\text{*}
\rangle\rangle\, C\, \langle\langle \epsilon\text{*}|\\ [15pt]
&=& \displaystyle \sum_{\mu,\nu\in E,E^{*}}\ket{\mu}C_{\mu\nu}^{*}\bra{\nu}=|\,\epsilon\rangle\rangle\, C\text{*}\, 
\langle\langle \epsilon|
\end{array}
\end{equation}
where we have arranged the matrix elements into the $N\times N$ matrices $C$ and $C\text{*}$.
The Overlap matrix $O$, the Hamiltonian $H$, and the expectation values of the angular momenta 
${\bf L},\,{\bf S}, {\bf J}$ are written in the direct basis, while the density matrix and the 
Green's function $G$ of the system are written in the dual basis, as will be explained below.
Matrix elements of operator products can be calculated using the above projectors
\begin{equation}
\bra{\mu}\hat{C}\,\hat{D}\ket{\nu}=\bra{\mu}\,\hat{C}\,|\,\epsilon\rangle\rangle 
\langle\langle\epsilon\text{*}\,|\,\epsilon\text{*}\rangle\rangle\,
\langle\langle \epsilon\,|\,\hat{D}\,\ket{\nu}=C_{\mu\epsilon}\,O^{-1}\,D_{\epsilon\nu}
\end{equation}
where $C_{\mu\epsilon}$ and $D_{\epsilon\nu}$ are $N$-dimensional row and column vectors of matrix elements, respectively.
The overlap matrix above simplifies if one of the operators is expressed in the dual basis. For example, the
product of an operator $\hat{C}$ times the Green's function $\hat{G}$ becomes
\begin{equation}
\bra{\mu}\,\hat{C}\,\hat{G}\,\ket{\nu\text{*}}=\langle\mu|\,\hat{C}\,|\epsilon\rangle\rangle\,\langle\langle\epsilon\text{*}|\,\hat{G}\,|\nu\text{*}\rangle=
C_{\mu\epsilon}\,G_{\epsilon\nu}^{*}
\end{equation} 
We address now how to define a local operator in terms of a subset A of the PAO basis. 
Any operator can be decomposed into four operator pieces
\begin{eqnarray}
\hat{C}
&=&\left(\hat{P}_A^\dagger+\hat{P}_B^\dagger\right)\hat{C}\left(\hat{P}_A+\hat{P}_B\right) \nonumber \\
&=&\underbrace{\hat{P}_A^\dagger\,\hat{C}\,\hat{P}_A}_{\hat{C}_{AA}}
+  \underbrace{\hat{P}_A^\dagger\,\hat{C}\,\hat{P}_B}_{\hat{C}_{AB}}
+  \underbrace{\hat{P}_B^\dagger\,\hat{C}\,\hat{P}_A}_{\hat{C}_{BA}}
+  \underbrace{\hat{P}_B^\dagger\,\hat{C}\,\hat{P}_B}_{\hat{C}_{BB}}\nonumber\\
&=& \begin{pmatrix}|\,A\text{*}\rangle\rangle\, |\,B\text{*}\rangle\rangle\end{pmatrix}
\begin{pmatrix}C_{AA}&C_{AB}\\C_{BA}&C_{BB}\end{pmatrix}
\begin{pmatrix} \langle\langle A\text{*}|\\\langle\langle B\text{*}|\end{pmatrix}.
\end{eqnarray}
$\hat{C}_{AA}$ above defines the {\it on-site operator} and the corresponding projection is called 
{\it On-site Projection} (OP). Soriano and Palacios\cite{Juanjo14} developed the above analysis, and 
described the different problems arising when trying to define local projections and operators.  

We propose here the following definition of a hermitian {\it local operator}:
\begin{eqnarray}  
\label{eqn:newprojection}
\hat{C}_A&=&\hat{C}_{AA}+\frac{1}{2}\,(\hat{C}_{AB}+\hat{C}_{BA}) \nonumber \\
&=& 
\begin{pmatrix}|\,A\text{*}\rangle\rangle\, |\,B\text{*}\rangle\rangle\end{pmatrix}
\begin{pmatrix}C_{AA}&\frac{1}{2}\,C_{AB}\\\frac{1}{2}\,C_{BA}&0\end{pmatrix}
\begin{pmatrix} \langle\langle A\text{*}|\\\langle\langle B\text{*}|\end{pmatrix}
\end{eqnarray}
and we will refer to the corresponding projection as {\it Local Projection} (LP). Notice that this operator fulfills already the most 
basic sum rule
\begin{equation}
\hat{C}=\hat{C}_A+\hat{C}_B
\end{equation}
We will show later in the article that this definition of a local operator fulfills two additional sum rules, 
while this is not the case for the on-site operator. 

\subsection{Spin degree of freedom}
We introduce now the Hilbert space of spin states of an electron ${\cal E}_S$, that we shall always express
in a basis of eigenstates of $\hat{S}_z$. Then the spin operator is 
\begin{equation}
\hat{\bf S}=\begin{pmatrix}\ket{\uparrow}\,\ket{\downarrow}\end{pmatrix}\,\frac{\hbar}{2}\,{\boldsymbol \tau}\,
\begin{pmatrix}\bra{\uparrow}\\\bra{\downarrow}\end{pmatrix}
\end{equation}
where ${\boldsymbol\tau}$ is the vector of Pauli matrices.
The electron spin degree of freedom is included in the PAO scheme by performing the direct product 
${\cal E}={\cal E}_N\otimes{\cal E}_S$. 
The spin operator is expressed in ${\cal E}$ as $\hat{\bf S}=\frac{\hbar}{2}\, \hat{I}\otimes {\boldsymbol \tau}$, and
in the PAO basis as
\begin{equation}
{\bf S} = \langle\langle\epsilon \,|\,\hat{\bf S}\,|\, \epsilon\rangle\rangle= 
\frac{\hbar}{2}\,O\otimes {\boldsymbol\tau} 
\end{equation} 
We shall drop from now on the direct product notation since there is no risk of confusion. Scalar and vector operators acting on ${\cal E}$ can be written as
\begin{equation}
\begin{array}{lcl}
\hat{A}_{\cal E}&=& \displaystyle \frac{1}{2}\,(\hat{A}_0\,\tau_0+\hat{\bf A}\cdot {\boldsymbol \tau})\\ [10pt]
\hat{\bf A}_{\cal E}&=& \displaystyle \frac{1}{2}\,(\hat{\bf A}\,\tau_0+\hat{A}_v\,{\boldsymbol \tau})
\end{array} 
\end{equation}
where $\hat{A}_0,\,\hat{A}_v,\,\hat{\bf A}$ act on ${\cal E}_N$ and  ${\cal E}_N^*$.

\subsection{Orbital angular momentum operator}
The orbital angular momentum with respect to a point ${\bf R}_A$ is 
\begin{equation}
{\hat{\bf L}}^A=(\hat{\bf r}-{\bf R}_A)\times\hat{\bf p}\,\,\tau_0
\end{equation}
We assume that ${\bf R}_A$ is centered at an atom $A$. We express this operator in the PAO basis as 
\begin{eqnarray}
\langle\langle\epsilon\, |\,\hat{\bf L}^A\, |\,\epsilon\rangle\rangle&=&
\begin{pmatrix}
\langle\langle A |\hat{\bf L}^A\,|\, A\rangle\rangle&\langle\langle A \,|\,\hat{\bf L}^A \,|\,B\rangle\rangle \nonumber \\ 
\langle\langle B \,|\,\hat{\bf L}^A\,|\, A \rangle\rangle&\langle\langle B\,|\, \hat{\bf L}^A \,|\,B\rangle\rangle
\end{pmatrix}\\ [5pt]
&=&\begin{pmatrix}O_{AA}\,{\bf {\cal L}}^A&{\bf {\cal L}}^A\,O_{AB}\\O_{BA}\,{\bf {\cal L}}^A&
{\bf L}^A_{BB}
\end{pmatrix}
\end{eqnarray}
where ${\bf {\cal L}}^A=\braket{L_A M_A|\hat{\bf L}^A| L_A M_A}$ is the angular momentum atomic matrix corresponding to atom $A$. 
Transforming from real $\ket{L M}$ to complex spherical harmonics $\ket{L m}$ is achieved by the matrix 
$R(L)$,
\begin{equation}
\ket{L M}=\sum_{m=-L}^L \braket{ L m | L M}\ket{L m}= \sum_{m=-L}^L R(L)_{m M}\ket{L m}
\end{equation}
 The matrix element ${\bf L}^A_{B B}=\langle\langle B\,|\, \hat{\bf L}^A\, |\,B\rangle\rangle$ can be determined using the following {\it Quantum Steiner Theorem}
 \begin{align}
& {\bf L}^A_{B B} =\langle\langle B\,|\,(\hat{\bf r}- {\bf R}_A)\times\hat{\bf p}\,|\,B\rangle\rangle \nonumber \\&=
\langle\langle B\,|\,(\hat{\bf r}- {\bf R}_B)\times\hat{\bf p}\,|\,B\rangle\rangle+
({\bf R}_A-{\bf R}_B)\times\langle\langle B\,|\,\hat{\bf p}\,|\, B\rangle\rangle\nonumber\\
 &={\bf {\cal L}}^B\,O_{B B}+({\bf R}_A-{\bf R}_B)\times\langle\langle B\,|\,\hat{\bf p}| \,B\rangle\rangle
 \end{align}
 Finally, the matrix elements of the linear momentum operator can be determined using standard methods 
 employed to determine e.g., kinetic energy terms\cite{siestapaper}. The above algebraic manipulations might 
 be useful to reduce the full Brillouin zone $k$-point sampling to smaller Brillouin zones by applying rotation-related 
 symmetry transformations.
 
\subsection{Variational eigenstates and eigenenergies}
We assume now that the physical system under consideration consists of a Bravais lattice having  $M$ unit cells,
defined by the Bravais vectors ${\bf R}_m$ with $m=1,\dots,\,M$. We choose a PAO basis set of $P$ orbitals within 
each unit cell ${\bf R}_m$ whose wave-functions are centered at the atomic positions ${\bf d}_\mu$ and
that are described by PAO kets $\ket{m,\mu}$ with $\mu=1,\dots,\,P$. We assume that the electron dynamics of the system is described by the Hamiltonian $\hat{H}$.  We then expand the Hamiltonian eigenstates in the variational basis:
\begin{eqnarray}
\ket{\psi_{{\bf k},n}}=\frac{1}{\sqrt{M}}\,\sum_{m,\mu}\,c_{{\bf k},n}(\mu)\,e^{i {\bf k}\cdot{\bf R}_m}\,
\ket{m,\mu}
\end{eqnarray}
where the eigenstate index is $n=1,\dots, P$. By collecting all PAOs belonging to the cell $m$ into the
vector of ket states $|\,{\bf R}_m\rangle\rangle=(\,\ket{m,1},\,\ket{m,2},\,\dots,\ket{m,P}\,)$, and rearranging 
the wave-function coefficients and PAOs into the Bloch vectors
\begin{eqnarray}
\underline{C_{{\bf k},n}}&=&
\begin{pmatrix}c_{{\bf k},n}(1)\\c_{{\bf k},n}(2)\\\dots\\c_{{\bf k},n}(P)\end{pmatrix}\\
\underline{\ket{\phi_{\bf k}}}&=&\frac{1}{\sqrt{M}}\,\sum_{m=1}^M \,e^{i {\bf k}\cdot{\bf R}_m}\,
\ket{{\bf R}_m}\rangle
\end{eqnarray}
the variational eigenfunctions can be rewritten as
\begin{eqnarray}
\ket{\psi_{{\bf k},n}}= \underline{\ket{\phi_{\bf k}}}\cdot \underline{C_{{\bf k},n}}
\end{eqnarray}
Notice that the wave-function coefficients $\underline{C_{{\bf k},n}}$ can be identified by taking the scalar product of 
$\ket{\psi_{{\bf k},n}}$ with the dual basis,
\begin{eqnarray}
\langle\braket{\epsilon\text{*}|\psi_{{\bf k},n}}=\frac{1}{\sqrt{M}}\,
\begin{pmatrix}e^{i {\bf k}\cdot{\bf R}_1}\\e^{i {\bf k}
\cdot{\bf R}_2}\\\dots\\e^{i {\bf k}\cdot{\bf R}_M}\end{pmatrix}\otimes\underline{C_{{\bf k},n}}
\end{eqnarray}
The variational eigenvalues and eigenvectors are obtained by minimizing the expectation value of Hamiltonian
\begin{eqnarray}
E_{{\bf k},n}=\frac{\braket{\psi_{{\bf k},n}|\,\hat{H}\,|\psi_{{\bf k},n}}}{\braket{\psi_{{\bf k},n}\,|\,
\psi_{{\bf k},n}}}=\frac{\underline{C_{{\bf k},n}}^{\dagger}\,H_{\bf k}\,\underline{C_{{\bf k},n}}}
{\underline{C_{{\bf k},n}}^{\dagger}\,O_{\bf k}\,\underline{C_{{\bf k},n}}}
\end{eqnarray}
where the momentum-space $P\times P$ Hamiltonian and overlap matrices are
\begin{align}
& H_{\bf k}=\underline{\bra{\phi_{\bf k}}}\hat{H}\underline{\ket{\phi_{\bf k}}}=\dfrac{1}{M}\sum_{{\bf R}_j}\, 
e^{-i{\bf k}\cdot({\bf R}_j-{\bf R}_0)}\,\langle\langle{\bf R}_j|\hat{H}|{{{\bf R}_0}}\rangle\rangle \\
& O_{\bf k}=\underline{\bra{\phi_{\bf k}}}\,\underline{\phi_{\bf k}\rangle}=
\dfrac{1}{M}\sum_{{\bf R}_j}\, e^{-i{\bf k}\cdot({\bf R}_j-{\bf R}_0)}\, \langle\langle{\bf R}_j\,|\,{\bf R}_0\rangle\rangle
\end{align}
Here $j$ runs over the supercell of unit cells coupled to the unit cell ${\bf R}_0$, that we take as reference.
The variational eigenvalues and eigenvectors are found by solving the conventional secular equation
\begin{eqnarray}
H_{\bf k}\,\underline{C_{{\bf k},n}}&=&E_{{\bf k},n}\,O_{\bf k}\,\underline{C_{{\bf k},n}}
\end{eqnarray}
The requirement that the eigenvector norm be positive, $\braket{\psi_{{\bf k},n}\,|\,
\psi_{{\bf k},n}}\,>\, 0$, means that $O_{\bf k}$ is a positive definite matrix. This in turn
implies that the eigenvalues $E_{{\bf k},n}$ are real and that the wave-function
coefficients are $O_{\bf k}$-orthogonal\cite{Parlett},
\begin{equation}
\underline{C_{{\bf k},n}}^\dagger\,O_{\bf k}\,\underline{C_{{\bf k},n'}}=\delta_{n,n'} 
\end{equation}
We gather now together the $P$ vectors $\underline{C_{{\bf k},n}}$ and build the $P\times P$ matrix
${\cal C}_{\bf k}=(\underline{C_{{\bf k},1}},\dots,\underline{C_{{\bf k},P}})$.
The secular equation and orthogonality relationships can be rewritten in a compact form as
\begin{equation}
\begin{aligned}
& H_{\bf k}\,{\cal C}_{\bf k}=\,O_k\,{\cal C}_{\bf k}\,H_{\bf k}^{\text{diag}}\\
& {\cal C}_{\bf k}^\dagger\,O_{\bf k}\,{\cal C}_{\bf k} = I_N \\
& H_k^{\text{diag}} = \text{diag}(E_{k,1},\dots,E_{k,P}) 
\end{aligned}
\end{equation}
so that 
\begin{equation}
{\cal C}_{\bf k}^\dagger\,H_{\bf k}\,{\cal C}_{\bf k} = H_{\bf k}^\text{diag} \, .
\end{equation}

\subsection{Green's function}
The retarded Green's function is defined in the usual one-body notation as
\begin{eqnarray}
\hat{G}(\omega)&=&\sum_{{\bf k},n}\,\frac{\ket{\psi_{{\bf k},n}}\bra{\psi_{{\bf k},n}}}{\omega-E_{{\bf k},n}
+i\,\delta}=\ket{\epsilon}\rangle\,G\text{*}(\omega)\,\langle\bra{\epsilon}\\
G\text{*}(\omega)&=&\langle\braket{\epsilon\text{*}|\hat{G}(\omega)|\epsilon\text{*}}\rangle=\sum_{{\bf k},n}\,
\frac{\langle\braket{\epsilon\text{*}| \psi_{{\bf k},n}}\braket{\psi_{{\bf k},n}| \epsilon\text{*}}\rangle}
{\omega-E_{{\bf k},n}+i\,\delta} \nonumber\\
&=&\sum_k\,D_{\bf k}\otimes G_{\bf k}^* (\omega)
\end{eqnarray}
where the supercell phases matrix $D_{\bf k}$ and unit-cell Green's function $G_{\bf k}^*(\omega)$ matrices are
\begin{eqnarray}
D_{\bf k} M&=&\,\begin{pmatrix}e^{i {\bf k}\cdot{\bf R}_1}\\\dots\\e^{i {\bf k}\cdot{\bf R}_M}\end{pmatrix}\begin{pmatrix}e^{-i {\bf k}\cdot{\bf R}_1},&
\dots,\,&e^{-i {\bf k}\cdot{\bf R}_M}\end{pmatrix}\\
G_{\bf k}^*(\omega)&=&{\cal C}_{\bf k}\,\left((\omega+i\,\delta)\,I_N-H_{\bf k}^\text{diag}\right)^{-1}\,{\cal C}^\dagger_{\bf k}
\end{eqnarray}
and $\delta$ is an infinitesimal positive number.

\subsection{Global and local symmetry transformations}
Let a finite global transformation be generated by the operator $\hat{U}_\theta=e^{-\,i\,\theta\,\hat{\cal G}}$ where 
$\hat{\cal G}$ is the transformation generator. Application of the above transformation to an observable $\hat{C}$ defines 
the deviation operator 
\begin{eqnarray}
\delta \hat{C}_{\theta}=\hat{U}_\theta\,\hat{C}\,\hat{U}_\theta^\dagger-\hat{C}=|\,\epsilon\text{*}\rangle\rangle\,\delta C_\theta\,\langle\langle\epsilon\text{*}\,|
\end{eqnarray}
Let ${\cal G}=\langle\langle\epsilon|\hat{\cal G}|\epsilon\rangle\rangle$ and $C=\langle\langle\epsilon|\hat{C}|\epsilon\rangle\rangle$ be the generator and an observable expressed in the PAO basis, respectively. Then we can write
\begin{align}
& \langle\langle\epsilon|\,\hat{U}_\theta\,|\epsilon\rangle\rangle = e^{-i\,\theta\,{\cal G}\,O^{-1}}\,O\\
& \delta C_\theta=\langle\langle\epsilon\, |\,\delta \hat{C}_\theta\,|\epsilon\rangle\rangle = e^{-i\,\theta\,{\cal G}\,O^{-1}}\,C\,e^{i\,O^{-1}\,\theta\,{\cal G}}-C \, .
\end{align}
If the transformation is infinitesimal, the variation of $C$ to second order is
\begin{align}
\delta C_{\delta\theta}\,&=\,\delta C^1\,\delta \theta+\delta C^2\,\delta\theta^2\\
\delta C^1 & =\,i\,\langle\langle\epsilon|\,[\,\hat{C},\hat{\cal G}\,]\,|\epsilon\rangle\rangle\,=\,i\,(C\,O^{-1}\,{\cal G}\,-\,{\cal G}\,O^{-1}\,C) \\
\delta C^2 & =\frac{1}{2}\,\langle\langle\epsilon|\,[\,[\,\hat{\cal G},\hat{C}\,],\,\hat{\cal G}\,]\,|\epsilon\rangle\rangle \nonumber\\
&= {\cal G}\,O^{-1}C\,O^{-1}{\cal G} \nonumber\\
&-\frac{1}{2}\,(C\,{\cal G}\,O^{-1}{\cal G}\,O^{-1}+{\cal G}\,O^{-1}{\cal G}\,O^{-1}C) \, .
\end{align}
The deviation operators of the global and the local transformations on a subset $A$ of the PAO basis can be written
as follows
\begin{eqnarray}
\label{eqn:local}
\delta \hat{C}_{\delta\theta}\,
&=& \begin{pmatrix}|\,A\text{*}\rangle\rangle\,|\,B\text{*}\rangle\rangle\end{pmatrix}
\begin{pmatrix}\delta C_{_{\delta\theta},AA}&\,\delta C_{_{\delta\theta},AB}\\\delta C_{_{\delta\theta},BA}&\delta C_{_{\delta\theta},BB}\end{pmatrix}
\begin{pmatrix}\langle\langle A\text{*}|\\ \langle\langle B\text{*}|\end{pmatrix} \\ [15pt]
\delta \hat{C}_{\delta\theta,A}\,
&=& \begin{pmatrix}|\,A\text{*}\rangle\rangle\,|\,B\text{*}\rangle\rangle\end{pmatrix}
\begin{pmatrix}\delta C_{_{\delta\theta},AA}&\frac{1}{2}\,\delta C_{_{\delta\theta},AB}\\\frac{1}{2}\,\delta C_{_{\delta\theta},BA}&0\end{pmatrix}
\begin{pmatrix}\langle\langle A\text{*}|\\ \langle\langle B\text{*}|\end{pmatrix}\nonumber\\
\end{eqnarray}

\subsection{Spin rotations}
A global spin rotation around a unit vector ${\bf u}$ is generated by the spin operator 
$\hat{\cal G}=\hat{{\bf S}}\cdot{\bf u}$ 
whose matrix elements are ${\cal G}=\frac{1}{2}\,O\,\tau^{\bf u}$, where the matrix 
$\tau^{\bf u}={\boldsymbol \tau}\cdot {\bf u}$. The presence of the overlap matrix in the definition of ${\cal G}$ 
allows us to
simplify considerably the algebra since the products ${\cal G}\,O^{-1}$ and $O^{-1}\,{\cal G}$ become just the matrix 
$T^{\bf u}=I_N\, \tau^{\bf u}$. As a consequence, the matrix elements of the deviation operator are
\begin{equation}
\begin{array}{lcl}
\delta C_{{\bf u},\delta\theta}\,&=&\,\delta C^1_{\bf u}\,\delta \theta+\delta C^2_{\bf u}\,\delta\theta^2\\ [5pt]
\delta C^1_{\bf u}&=&\, \displaystyle \frac{i}{2}\,[\,C,T^{\bf u}\,] \\ [7pt]
\delta C^2_{\bf u}&=& \displaystyle \frac{1}{8}\,[\,[\,T^{\bf u},C\,],T^{\bf u}\,]
\end{array}
\end{equation}

\section{Perturbative analysis of the classical Hamiltonian}
\label{section:Perturbativeclassical}
We start this section by rewriting the classical Hamiltonian in Eq. (\ref{eqn:Heisenberg2}) in a more
convenient way. We first subtract the constant term $\sum_i K_i^{zz}\,I_3$, 
and redefine the intra-atomic anisotropy matrices as
\begin{eqnarray}
\tilde{K}_i=K_i-K_i^{zz}\,I_3=\begin{pmatrix} K_i^{xx}-K_i^{zz} & K_i^{xy} &  K_i^{xz} \\  K_i^{yx} 
& K_i^{yy}-K_i^{zz} & K_i^{yz} \\ K_i^{zx} &  K_i^{zy} & 0 \end{pmatrix}\nonumber\\
\end{eqnarray}
This renormalized anisotropy matrix has only five independent matrix elements.
We then introduce the symmetric and antisymmetric parts of the
exchange tensor $J^{s,a}=(J\pm J^T)/2$ that enable us to define the Dzyaloshinskii-Morilla (DM) and the symmetric vectors,
\begin{eqnarray}
{\bf D}_{ij}&=&(J_{ij}^{yz,a},J_{ij}^{zx,a},J_{ij}^{xy,a})\\ 
{\bf S}_{ij}&=&(J_{ij}^{yz,s},J_{ij}^{zx,s},J_{ij}^{xy,s})
\end{eqnarray}
such that
\begin{eqnarray}
\label{Js+Ja}
J=J^s+J^a&=&\begin{pmatrix} J^{xx} & S^z &  S^y \\  S^z & J^{yy} & S^x \\ S^y &  S^x & J^{zz} \end{pmatrix}+
\begin{pmatrix}   0    & D^z & -D^y \\ -D^z &   0    & D^x \\ D^y & -D^x &    0   \end{pmatrix}\nonumber\\
\end{eqnarray}
where the $i$, $j$ site subindices have been omitted for simplicity. We define now the Heisenberg exchange
constant  and the renormalized symmetric tensor, 
\begin{eqnarray}
J^H_{ij}&=&\Tr J^s_{ij}/3\\
J^S_{ij}&=&J^s_{ij}-J^H_{ij}\,I_3 \, ,
\end{eqnarray}
where $I_3$ is the
identity $3\times 3$ matrix. As a consequence, the classical Hamiltonian in Eq. (\ref{eqn:Heisenberg2})
can be rewritten as
\begin{eqnarray}
\label{eqn:Heisenberg3}
H&=&\,\frac{1}{2}\,\sum_{i\neq j}\,J_{ij}^H\,{\bf e}_i\cdot{\bf e}_j+
\frac{1}{2}\,\sum_{i\neq j}\,{\bf e}_i\,J_{ij}^S\,{\bf e}_j \nonumber \\
&&+\,\frac{1}{2}\,\sum_{i\neq j}\,{\bf D}_{ij}\cdot({\bf e}_i\times{\bf e}_j)+
\sum_i\,{\bf e}_i\,\tilde{K}_i\,{\bf e}_i \, .
\end{eqnarray}
The perturbative analysis \cite{LKAG}} assumes that Hamiltonian (\ref{eqn:Heisenberg3}) has a 
local or global energy minimum when the unit vectors ${\bf e}_i$ are all collinear 
(either aligned ferromagnetically FM or antiferromagnetically AFM) to 
each other along a direction given by the unit vector ${\bf o}$, $E_G=H(\{{\bf o}\})$. 
Then, performing an infinitesimal rotation of the ${\bf o}$ vector at site $A$, of magnitude 
$\delta \theta$ around an axis defined by the unit vector ${\bf u}_A$, delivers the rotated vector
${\bf e}_A$ that is given at second order in $\delta \theta$ by
\begin{equation}
\label{derivativesvectors}
\begin{array}{lcl}
{\bf e}_A&\simeq&{\bf o}\,+{\bf o}_A^1\,\delta\theta\,+\,{\bf o}_A^2\,\delta \theta^2 \\ [5pt]
{\bf o}_A^1 &=& {\bf u}_A\times{\bf o} \\ [5pt]
{\bf o}_A^2 &=&\frac{1}{2}({\bf u}_A\,({\bf u}_A \cdot {\bf o})-{\bf o}) \, .
\end{array} 
\end{equation}
The energy variation  expanded to second order is
\begin{widetext}
\begin{eqnarray}
 \label{firstandsecondorder}
E(\{{\bf o}\},\,{\bf u}_A)\,&\simeq&\, E_0\,+\,E_A^1\,\delta \theta + E_A^2\,\delta \theta^2\\
E_A^1(\{{\bf o}\},\,{\bf u}_A)\,&=&\frac{1}{2}\left({\bf o}\,\left(\,\sum_i J_{iA}+2\,K_A\,\right)\,
 {\bf o}_A^1+\,{\bf o}_A^1\,\left(\,\sum_i J_{Ai}+2\,K_A\,\right)\, {\bf o}\right) \\
E_A^2(\{{\bf o}\},\,{\bf u}_A)\,&=&\frac{1}{2}\,\left(\,{\bf o}\,\left(\,\sum_i J_{iA}+2\,K_A\,\right)\,
 {\bf o}_A^2+ {\bf o}_A^2\,\left(\sum_i\,J_{Ai}+2\,K_A\,\right)\,{\bf o}\right)\,+\,{\bf o}_A^1\,K_A\,{\bf o}_A^1 \, ,
\end{eqnarray}
\end{widetext}
where $E^1_A=0$ according to our assumption that the Hamiltonian (\ref{eqn:Heisenberg3}) has a local 
or global 
extremum at the configuration where the spins are all aligned along ${\bf o}$. The above equations are 
completely general.

We describe now how all the matrix elements in each $J_{ij}$ and $\tilde K_i$ tensor can be obtained
performing rotations about  two axes perpendicular to ${\bf o}$. 
We denote these two axes by ${\bf v}$ and ${\bf w}={\bf o}\times{\bf v}$. The vectors
${\bf o},\,{\bf v},\,{\bf w}$ can then be chosen to be the three cartesian axes ${\bf x},\,{\bf y},\,{\bf z}$ of the coordinate system  that define the Hamiltonian 
(\ref{eqn:Heisenberg3}), or any cyclic permutation of them.

The fact that the first-order
variations are zero leads to the following two sum rules:
\begin{equation}
\label{eqn:firstordersumrules}
\begin{array}{lcl}
\sum_i\,J_{iA}^{o v}\,+2\,K_{A}^{o v}=0\\ [5pt]
\sum_i\,J_{iA}^{o w}\,+2\,K_{A}^{o w}=0 \, ,
\end{array}
\end{equation}
where we have taken advantage of the fact that $J_{iA}=J_{Ai}^T$ and $K_i=K_i^T$.
These equations allow us to determine two of the three off-diagonal matrix elements of the $\tilde{K}_i$-matrices provided 
that the
related exchange matrix elements are known.

Second-order energy variations give access to the two diagonal elements of the anisotropy matrix,
\begin{equation}
\label{eqn:singlesite}
\begin{array}{lcl}
K_A^{ww}-K_A^{vv}&=&E^2_A(\{{\bf o}\},{\bf v})-E^2_A(\{{\bf o}\},{\bf w})\\ [5pt]
K_A^{ww}-K_A^{oo}&=&E^2_A(\{{\bf v}\},{\bf o})-E^2_A(\{{\bf v}\},{\bf w})\\ [5pt]
K_A^{vv}-K_A^{oo}&=&E^2_A(\{{\bf w}\},{\bf o})-E^2_A(\{{\bf w}\},{\bf v}) \, .
\end{array}
\end{equation}
This set of six equations allows us to perform consistency checks on the two diagonal matrix elements.

The remaining off-diagonal matrix element $K_A^{vw}$ can be obtained by computing second-order energy variations
about an axis defined by a unit vector lying at the diagonal of ${\bf v}$ and ${\bf w}$,
\begin{equation}
K_A^{vw}=\frac{1}{2}\left(E^2_A(\{{\bf o}\},{\bf v})+E^2_A(\{{\bf o}\},{\bf w})\right)-E^2_A\left(\{{\bf o}\},\frac{{\bf v}+{\bf w}}{\sqrt{2}}\right) \, . 
\end{equation}

Furthermore, for those
cases where the Hamiltonian (\ref{eqn:Heisenberg3}) has extrema along all the three  
directions ${\bf o}$, ${\bf v}$ and ${\bf w}$, the conditions \eqref{eqn:firstordersumrules} can be recast as the following set of six sum rules,
\begin{align}
\label{eqn:DMsumrule}
& \sum_i {\bf D}_{iA}=0 \\
\label{eqn:JKsumrule}
& \sum_i J_{iA}^{S,\alpha\beta}+2\,K_A^{\alpha\beta}=0\,\,\,(\alpha,\beta \in \{o,v,w\},\,\alpha\ne \beta) \, . 
\end{align}
The first set of sum rules dictates that the sum of the DM vectors is identically zero. The second set of sum rules provide us with an alternative
way to determine all
three off-diagonal matrix elements of the $\tilde{K}_A$-matrix.
A third strategy to determine matrix elements of the $\tilde{K}_A$ matrix relies on taking advantage of the symmetries of the system under study. We will give an example in section \ref{subsection:supportedtrimer} below.

The exchange tensor $J_{AB}$ for a given site pair $(A,\,B)$ can be calculated by performing
simultaneous rotations at the two sites, along the directions ${\bf u}_A$,
${\bf u}_B$ with the same magnitude $\delta \theta$. A straightforward calculations yields
\begin{eqnarray}
E(\{{\bf o}\},\,{\bf u}_A,\,{\bf u}_B)&\simeq& E_0+E_A^2(\{{\bf o}\},\,{\bf u}_A)+
E_B^2(\{{\bf o}\},\,{\bf u}_B)\nonumber\\
&&+E_\mathrm{int}(\{{\bf o}\},\,{\bf u}_A,\,{\bf u}_B)
\end{eqnarray}
where
\begin{equation}
E_\mathrm{int}(\{{\bf o}\},\,{\bf u}_A,\,{\bf u}_B)=\dfrac{1}{2}\left({\bf o}_A^1\,J_{AB}\,{\bf o}_B^1+{\bf o}_B^1\,J_{BA}\,{\bf o}_A^1\right)\,\delta\theta^2 \, ,
\end{equation}

The explicit expressions for the exchange tensor of a generic magnetic material can then be found
by choosing ${\bf o}$  to lie subsequently along each of the three coordinate axes while the unit vectors
${\bf u}_A$, ${\bf u}_B$ are chosen along the other two perpendicular axes  ${\bf v}$ and ${\bf w}$. We find
\begin{flalign}
\label{eqn:doublesite}
& \delta\theta^2\,J_{AB}^{ww}=E_\mathrm{int}(\{{\bf o}\},\,{\bf v},\,{\bf v}) & \nonumber \\
& \delta\theta^2\,J_{AB}^{S,vw}=-\frac{1}{2}\left(E_\mathrm{int}(\{{\bf o}\},\,{\bf v},\,{\bf w}) + E_\mathrm{int}(\{{\bf o}\},\,{\bf w},\,{\bf v}) \right) & \\
&   \delta\theta^2\,D_{AB}^{\bf o}=\dfrac{\epsilon_{o v w}}{2}\left(E_\mathrm{int}(\{{\bf o}\},\,{\bf v},\,{\bf w}) - E_\mathrm{int}(\{{\bf o}\},\,{\bf w},\,{\bf v}) \right) \, , & \nonumber
\end{flalign}
where $\epsilon_{ovw}$ denotes the Levy-Civita tensor.
These equations allow us to determine all the matrix elements of the symmetric tensors  $J^S_{AB}$ and the DM
vectors $\mathrm{\bf D}_{AB}$. Furthermore, they provide a means to test
the accuracy of any numerical implementation because the number of equations is larger than the number of unknowns.

\begin{figure}[ht]
    \centering
    \includegraphics[width=\columnwidth]{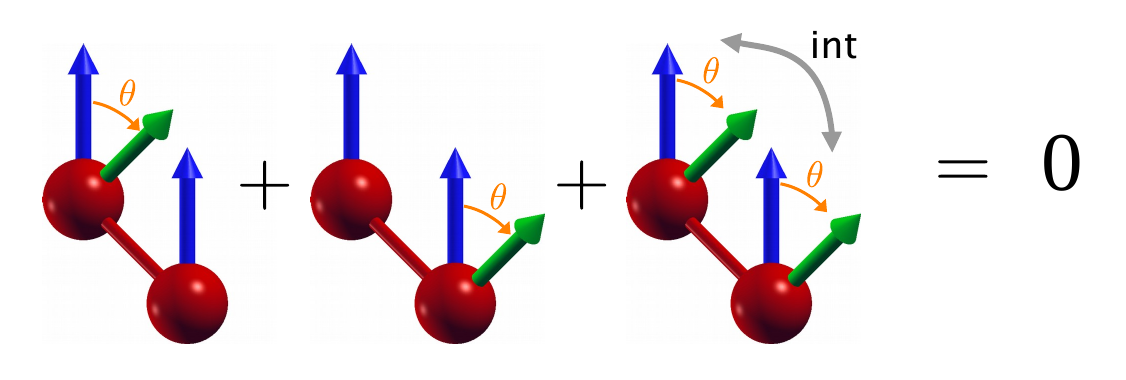}
    \caption{Sum rule for the global spin rotation of a dimer system: the sum of the energy of the local spin 
    rotation of both atoms plus the interaction energy is zero in the absence of spin-orbit interaction.}
    \label{fig:figure1}
\end{figure}

The above analysis gives another stringent sum rule for the case of a dimer system: if the spin-orbit
interaction is absent, then the dimer has rotational symmetry, so a global spin rotation by the
vector ${\bf u}$ does not change the energy. Therefore 
\begin{eqnarray}
\label{eqn:dimersumrule}
E_A^2(\{{\bf o}\},\,{\bf u})+E_B^2(\{{\bf o}\},\,{\bf u})
+E_\mathrm{int}(\{{\bf o}\},\,{\bf u},\,{\bf u})=0\nonumber\\
\end{eqnarray}
A graphical explanation of the dimer sum rule is shown in Fig. \ref{fig:figure1}.
Furthermore, even if the spin-orbit interaction is included, a dimer still has rotational symmetry along
the dimer axis, so a global spin rotation around that symmetry axis also yields the above sum rule. 

We finally note that the choice perfomed above of the orthogonal axes can be generalized by selecting 
non-orthogonal rotation directions as discussed in detail  in Ref. [\onlinecite{Udvardi_relativistic_Jij}]. 
This latter choice has the benefit that it can exploit symmetries of 
the system being investigated that are not compatible with the choice of orthogonal axes. 

\section{Perturbative analysis of the electron Hamiltonian}
\label{section:KS}
A number of physical issues and technicalities arise when trying to establish the 
equivalent perturbative analysis of the electron Hamiltonian, that must be handled correctly to avoid 
nonphysical results or inaccuracies.  We write down in this section a complete roadmap 
including a brief discussion of some well known issues.  The first subsection shows that energy variations
upon an infinitesimal transformation can be determined using the Kohn-Sham Hamiltonian of the unperturbed system.
The second subsection explains how to extract the different pieces of the KS Hamiltonian. The third subsection describes how to perform correctly rotations of the angular momentum in the KS Hamiltonian. It also addresses
the identification of the correct angular momentum operator in the DFT2S mapping.
The fourth subsection establishes the LKAG formula for a PAO basis set. The final subsection summarizes the
steps to be taken to perform the DFT2S mapping. 

\subsection{Magnetic Force Theorem}
The torque-based DFT2S mapping consists of determining the energy variation of the magnetic material 
upon application of an infinitesimal perturbation on its many-body electron Hamiltonian. A difficulty arises
because this energy variation is calculated from the associated KS Hamiltonian. Let us assume that
we apply a perturbation $\delta V$ to the external potential. Let us also assume that the ground-state density of the unperturbed
and perturbed systems are $n^0$ and $n=n^0+\delta n$.  Then the KS Hamiltonians are
\begin{equation}
\label{eqn:hams}
\begin{array}{lcl}
H^0[n^0]&=&T+V_{HXC}[n^0]\\ [5pt]
H^F[n^0]&=&H^0[n^0]+\delta V \\ [5pt]
H[n]&=&T+V_{HXC}[n]+\delta V \\ [5pt]
&=&H^F[n^0]+\delta V_{HXC}[n^0] \, .
\end{array}
\end{equation}
In other words, the perturbed KS Hamiltonian is not simply $H^F$ as would be the case of a simple one-body
electron Hamiltonian, because the density $n$ changes, which yields a change $\Delta V_{HXC}[n^0]$ in the Hartree-Exchange-Correlation
fictitious potential $V_{HXC}$. Let the eigenenergies of the three Hamiltonians in Eq. (\ref{eqn:hams}) be 
\begin{equation}
\begin{array}{lcl}  
\epsilon_\alpha^0[n^0] &&\\ [5pt]
\epsilon_\alpha^F[n^0]&=&\epsilon_\alpha^0[n^0]+\delta\epsilon_\alpha^0[n^0]\\ [5pt]
\epsilon_\alpha[n]&=&\epsilon^F_\alpha[n^0]+\delta \epsilon_\alpha^F[n^0] \, , 
\end{array}
\end{equation}
respectively.
The ground 
state energy is not just given by the sum of eigenenergies, but double-counting terms 
must be subtracted
\begin{align}
& E^0[n^0]=\sum_\alpha \epsilon^0_\alpha[n^0]\,f_\alpha[n^0]-E_\mathrm{dc}[n^0]\\
& E[n]=\sum_\alpha \epsilon_\alpha[n]\,f_\alpha[n]-E_\mathrm{dc}[n]\nonumber\\
&=\sum_\alpha (\epsilon_\alpha^F[n^0]+\delta \epsilon_\alpha^F[n^0])\,f_\alpha[n^0]-E_\mathrm{dc}[n^0]-
\delta E_\mathrm{dc}[n^0] \, ,
\end{align}
where we assume that the continuity principle $f_\alpha[n]=f_\alpha[n^0]$ typical of quasi-particle Fermi Liquid
Theory is fulfilled. The magnetic force theorem \cite{MFT} states that potential relaxation effects and 
double-counting terms cancel each other,
 \begin{eqnarray}
\sum_\alpha\delta \epsilon_\alpha^F[n^0]\,f_\alpha[n^0]=\delta E_\mathrm{dc}[n^0] \, ,
 \end{eqnarray}
such that the energy variation can just be computed from the eigenenergies of $H^F$ just as if the many-body problem 
 was in effect a one-body problem,
 \begin{eqnarray}
 E[n] =E^0[n^0]+\delta E[n^0] =E^0[n^0]+ \sum_\alpha \delta\epsilon_\alpha^0[n^0]\,f_\alpha[n^0] \, . \nonumber \\
 \end{eqnarray}
 
\subsection{Extracting the different components of the KS Hamiltonian}
Within non-collinear DFT, the Exchange-Correlation potential is expanded as 
\begin{eqnarray}
\frac{1}{2}(V_{XC,0}\,\tau_0+{\bf V}_{XC}\cdot {\boldsymbol \tau})=
\frac{1}{2}(V_{XC,0}\,\tau_0+V_{XC}\,\hat{\bf m}\cdot {\boldsymbol \tau}) \, , \nonumber \\
\end{eqnarray} 
where the exchange field ${\bf V}_{XC}$ is assumed to be collinear  to the magnetization direction $\hat{\bf m}$ everywhere.
The KS Hamiltonian can then be written as
\begin{eqnarray}
H&=&\frac{1}{2}\,\,\left(H_0\,\tau_0+{\bf V}_\text{SO}\cdot{\boldsymbol \tau}+
{\bf V}_{XC}\cdot{\boldsymbol \tau}\right)
\end{eqnarray}
The scalar part of the Hamiltonian $H_0$ is comprised of the kinetic 
energy $T$, the lattice potential $V_L$, the Hartree potential  $V_H$, and the scalar part of the 
exchange-correlation potential $V_{XC,\text{0}}$, and ${\bf V}_\text{SO}$ is a sum of atomic spin-orbit potentials,
\begin{equation}
\begin{array}{lcl}
  H_0&=&T+V_L+V_H+V_{XC}^0 \\ [5pt]
  {\bf V}_\text{SO}&=& \displaystyle \sum_a\,V_\text{SO}(|{\bf r}-{\bf R}_a|)\,\,{\bf l}_a \, ,
\end{array}
\end{equation}
where the angular momentum operator with respect to a point ${\bf R}_a$ is ${\bf L}_a=\hbar\,{\bf l}_a$. 

We address now how to extract ${\bf V}_{XC}$ from the KS Hamiltonian. We define  the time-reversal operator 
$\hat{\cal T}=i\,\tau_2\,\hat{\cal C}$, where $\hat{\cal C}$ denotes charge conjugation and acts on ${\cal E}_N$. Application of ${\cal T}$ on $n({\bf r})$ leaves $n_0$ invariant 
and switches the sign of ${\bf m}$. Application of ${\cal T}$ on the KS Hamiltonian yields $H^\mathrm{TR}={\cal T}\,H\,{\cal T}^{-1}$. 
Then $H$ can be split into time-reversal-symmetric (TRS) and time-reversal-broken (TRB) parts
\begin{equation}
\label{eqn:htrb}    
\begin{array}{lcl}
H^\mathrm{TRS}&=& \displaystyle \frac{H+H^\mathrm{TR}}{2}= \displaystyle \frac{1}{2}\,(H_0\,\tau_0+{\bf V}_\text{SO}\cdot{\boldsymbol \tau})\\ [9pt]
H^\mathrm{TRB}&=& \displaystyle \frac{H-H^\mathrm{TR}}{2}= \displaystyle \frac{1}{2}\,{\bf V}_{XC}\cdot {\boldsymbol \tau} \, .
\end{array}
\end{equation}
Finally, $H_0$ and the vectors ${\bf V}_{XC}$ and ${\bf H}_\text{SO}$ can be found by taking the spin trace 
\begin{equation}
\begin{array}{rcl}
H_0&=& \Tr_S (H^\mathrm{TRS})\\ [5pt]
{\bf V}_\text{SO}&=& \Tr_S (H^\mathrm{TRS}\,{\boldsymbol \tau})\\ [5pt]
{\bf V}_{XC}&=& \Tr_S (H^\mathrm{TRB}\,{\boldsymbol \tau} )\, .
\end{array}
\end{equation}

\subsection{Rotating non-collinear Kohn-Sham Hamiltonians}
\label{sec:rotating}
We devote this section to analyse how to establish a faithful mapping between the KS and the classical
Hammiltonian (\ref{eqn:Heisenberg3}). Note that the spin-orbit interaction enters the classical Hamiltonian in two 
ways. First, the exchange constants $J_{AB}$ become $3\times 3$ matrices instead of scalars, and the 
intra-atomic anisotropy tensors $K_A$ become non-zero; second, the vectors ${\bf e}_A$ can now refer 
not only to the atomic spin, but can also represent the atomic total angular momentum.

\begin{figure}[ht]
\centering
\includegraphics[width=\columnwidth]{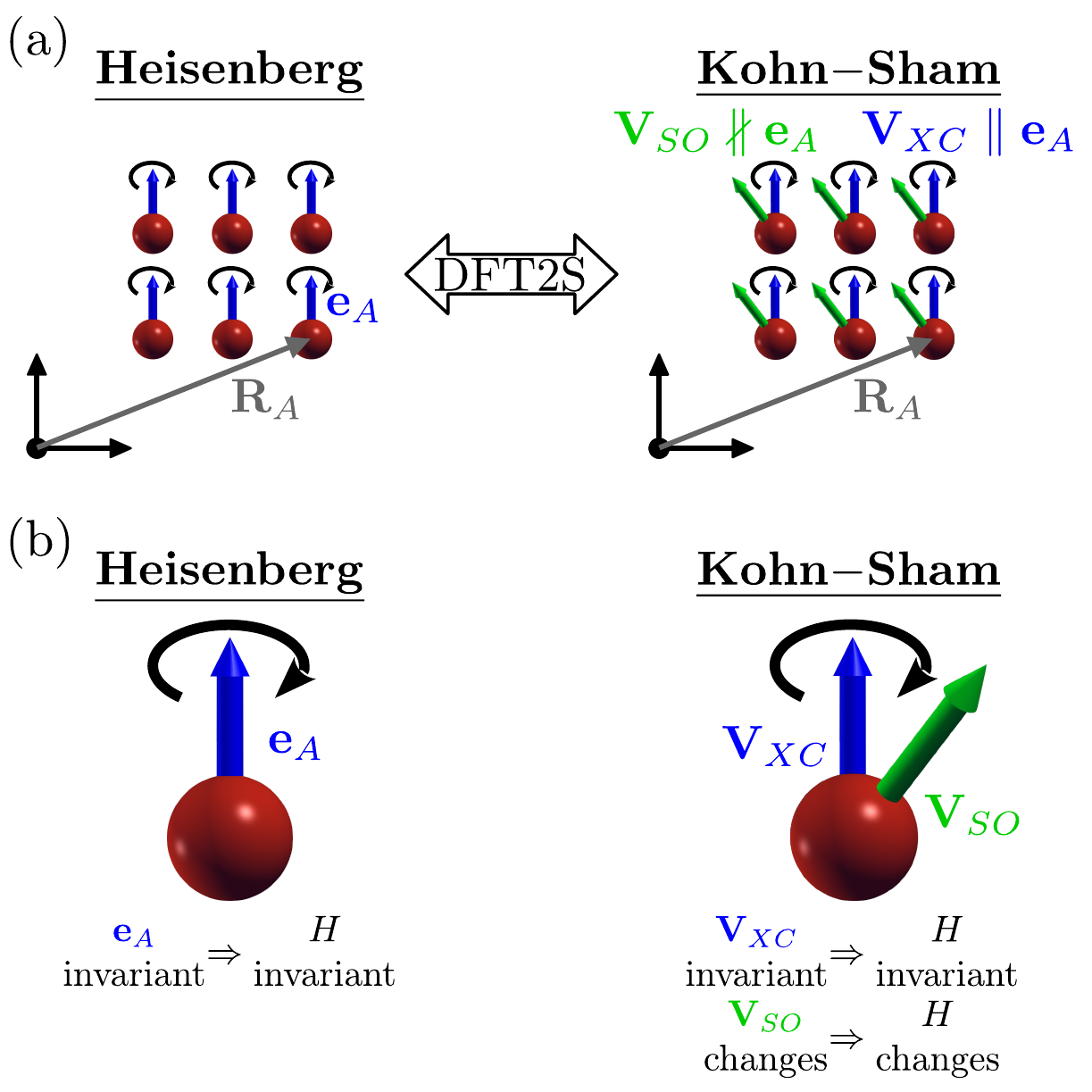}
\caption{Panel (a) illustrates the vectors appearing in the  classical (left) and
the KS Hamiltonian (right). It is also indicated by curved arrows which of those should be transformed upon rotations in order to achieve a 
faithful DFT2S mapping. ${\bf V}_{XC}$ is parallel to the atomic magnetization ${\bf e}_A$ by definition, while, quite generally, ${\bf V}_{SO}$ is not. Panel (b) shows graphically the impact of a rotation about the quantization axis on the classical Hamiltonian (left) 
and the KS Hamiltonian (right). The classical Hamiltonian is left invariant, while the KS Hamiltonian changes 
if ${\bf V}_{SO}$ is rotated, thus invalidating the DFT2S mapping. We will show in section \ref{IsolatedPtdimer} 
and table \ref{tabledimerrotation} how this reasoning is applied in the case of a platinum dimer.
}
    \label{Vxcfigure}
\end{figure}

We stress that in section \ref{section:Perturbativeclassical} we have performed  rotations 
over the classical Hamiltonian that operated only on the unit vectors ${\bf e}_A$. 
The DFT2S mapping is hence accomplished by performing transformations of the electron KS Hamiltonian where only 
the angular momentum vector at a given atom $A$ is rotated, while keeping all other vectors 
untouched. Possible angular momentum vectors are the atomic spin ${\bf m}_A$ and the atomic total angular 
momentum ${\bf J}_A$, and we 
identify ways to perform transformations of the KS 
Hamiltonian  where only the atomic angular momenta are rotated. We stress that other vectors appearing in 
the KS Hamiltonian such as ${\bf r}$ and ${\bf V}_\mathrm{SO}$ should remain invariant under the transformations.
This aspect of the DFT2S mapping is displayed schematically in Fig. \ref{Vxcfigure}. 

First we address rotations of the atomic spins in terms of the orientation of the local KS spin magnetization 
given the unit vector $\hat{\bf m}({\bf r})$ being a basic quantity of the Local Spin Density Approximation (LSDA) to DFT. We notice that 
$\hat{\bf m}({\bf r})$  enters the KS Hamiltonian only in the combination 
$V_{XC}({\bf r})\,\hat{\bf m}({\bf r})\cdot {\boldsymbol \tau}$.
We then remind that rotating the three-dimensional vector $\hat{\bf m}$ in the scalar product
by an angle $\theta$ around an axis defined by the unit vector ${\bf u}$ is equivalent to 
applying a reversed rotation on the $2\times 2$ Pauli
matrices because
\begin{equation} 
(\mathrm{R}\,\hat{\bf m})\cdot{\boldsymbol\tau}=
\hat{\bf m}\cdot \left(\mathcal{R}_S^{-1}\,{\boldsymbol \tau}\,\mathcal{R}_S\right)
\end{equation}
where $\mathrm{R}=\mathrm{R}(\theta,{\bf u})=e^{\theta\,{\bf u}\times}$ is a $3\times 3$ rotation matrix
according to Rodrigues' formula ($\times$ denotes the cross product), and the Pauli matrices are rotated
by the spin rotation matrix
\begin{eqnarray}
{\cal R}_S&=&{\cal R}_S(\theta,{\bf u})=e^{-i \theta {\bf S}\cdot {\bf u}/\hbar}=e^{-i \theta {\boldsymbol\tau}\cdot {\bf u}/2}
\end{eqnarray}
We apply a back-rotation to the different parts of the KS Hamiltonian and find

\begin{flalign}
& {\cal R}_S^{-1}\,H_0({\bf r})\,\tau_0\,{\cal R}_S = H_0({\bf r})\,\tau_0 \\ 
& {\cal R}_S^{-1} \left( {\bf V}_{SO}({\bf r})\cdot {\boldsymbol \tau} \right) {\cal R}_S =
{\bf V}_{SO}({\bf r})\cdot\left(\,{\cal R}_S^{-1}\,{\boldsymbol \tau}\,{\cal R}_S\right) \nonumber \\ 
& \hspace{3.2cm} = \left(\,\mathrm{R}\,{\bf V}_{SO}({\bf r})\,\right) \cdot{\boldsymbol \tau} \\ 
& {\cal R}_S^{-1} \left( {\bf V}_{XC}({\bf r})\cdot {\boldsymbol\tau} \right) {\cal R}_S = 
{\bf V}_{XC}({\bf r})\cdot (\,{\cal R}_S^{-1}\,{\boldsymbol\tau}\,{\cal R}_S\,)  \nonumber \\  
& \hspace{3.3cm} = \left(\,\mathrm{R}\,{\bf V}_{XC}({\bf r})\,\right)\cdot{\boldsymbol\tau} \, . 
\end{flalign}

This important result means that performing a spin rotation on the full Hamiltonian not only 
rotates ${\bf V}_{XC}$ and hence $\hat{\bf m}$. It also rotates the vector ${\bf V}_{SO}$, which is undesirable
if one wishes to accomplish a DFT2S mapping, as explained in the second paragraph of this section and in 
Figure \ref{Vxcfigure}.

We address now whether a DFT2S mapping can be achieved when the vector ${\bf e}_A$ refers to  the 
atomic total angular momentum at at atomic site $A$. We must therefore rotate the KS Hamiltonian 
using the total momentum ${\bf J}_A={\bf L}_A+{\bf S}$. This generates rotations whose 
operator is ${\cal R}_{J,A}\,= {\cal R}_{L,A}\,{\cal R}_S$.
Rotating the different parts of the KS Hamiltonian yields

\begin{flalign}
& {\cal R}_{J,A}^{-1}\,H_0({\bf r})\,\tau_0\,{\cal R}_{J,A} = H_0( \mathrm{R}_A^{-1} {\bf r})\,\tau_0 \\ 
& {\cal R}_{J,A}^{-1} \left( {\bf V}_{SO}({\bf r})\cdot {\boldsymbol \tau} \right) {\cal R}_{J,A}  
 = \left( \mathrm{R}_A^{-1}{\bf V}_{SO}(\mathrm{R}_A^{-1} {\bf r}) \right) \cdot \left( {\cal R}_S^{-1}\,{\boldsymbol \tau}\,{\cal R}_S\right) \nonumber \\ 
& \hspace{3.5cm} = {\bf V}_{SO}(\mathrm{R}_A^{-1} {\bf r}) \cdot {\boldsymbol \tau} \\
& {\cal R}_{J,A}^{-1} \left( {\bf V}_{XC}({\bf r})\cdot {\boldsymbol\tau} \right) {\cal R}_{J,A} = 
{\bf V}_{XC}(\mathrm{R}_A^{-1} {\bf r}) \cdot \left( {\cal R}_S^{-1}\,{\boldsymbol\tau}\,{\cal R}_S \right)  \nonumber \\  
& \hspace{3.6cm} = \left( \mathrm{R}_A {\bf V}_{XC}(\mathrm{R}_A^{-1} {\bf r}) \right) \cdot {\boldsymbol\tau} \, ,
\end{flalign}

which means that this transformation rotates not only the magnetization but also 
causes the space position ${\bf r}$ to rotate which is undesirable within the DFT2S mapping. 

All in all, a correct DFT2S mapping is valid if only the exchange field ${\bf V}_{XC}$ is 
spin-rotated. We will present several tests in section \ref{section:Tests} that demonstrate numerically this
conclusion. We should note however that this restriction does not hold when the Atomic Sphere Approximation 
is implemented. In this case the potentials and fields have a spherical shape 

so that the total angular momentum rotation of the Hamiltonian can be used to achieve the rotation of the exchange field only \cite{Udvardi_relativistic_Jij}.

We remind now that we have shown in section \ref{section:Perturbativeclassical} that determining {\em all} the 
tensors' components for a chosen cartesian coordinate system requires three different classical Hamiltonians, each 
of them having a uniform magnetic direction ${\bf o}$ oriented along one of the three cartesian axes.
We therefore need to compute Kohn-Sham Hamiltonians $H^{\bf m}$ where $\hat{\bf m}$ is oriented
along the three coordinate axes $x,\,y$ and $z$. These three Hamiltonians can be obtained by performing three 
different collinear DFT simulations.  We have
found that, especially for isolated
molecules and nanostructures, the three simulations lead to different magnetic multiplets making the application of the magnetic force theorem questionable. As a simpler approach we performed a single self-consistent DFT collinear simulation
where $\hat{\bf m}$ is oriented along any of the three axes resulting in the Hamiltonian $H^{\bf m}$. 
The exchange field  ${\bf V}_{XC}^{\bf m}$ is then extracted and rotated to deliver two new exchange fields 
${\bf V}_{XC}^i$ that are aligned  along  the other two coordinate axes. The two required KS 
Hamiltonians can then be obtained using $H^i=H^{TRS}+H^{i,\,TRB}$.

\subsection{The LKAG torque method}
The LKAG torque method to perform the DFT2S mapping consists of performing infinitesimal rotations on localized
angular moments of the Kohn-Sham fictitious system associated with a many-body Hamiltonian, computing the 
second-order energy variations of the KS system due to those rotations and equating them with the equivalent
classical expression given in Eq. (\ref{eqn:singlesite}) and (\ref{eqn:doublesite}). 

The magnetic force theorem demonstrates that the energy variations can be computed directly from the KS Hamiltonian
as if it represented a one-body system, because potential relaxation effects cancel variation in double counting 
terms.
Then, these energy variations of the one-body KS Hamiltonian can be determined using a variety of techniques, and
we use here the popular Lloyd's approach \cite{Zeller_Lloyd_formula}. We assume that we have a one-body Hamiltonian
$\hat{H}=\hat{H}_0+\hat{V}_A$ where $\hat{V}_A$ is a perturbation. Let $\hat{G}_0(\epsilon)$ and 
$\hat{G}(\epsilon)$ be the Green's functions associated to $\hat{H}_0$ and $\hat{H}$. Then the density of states is
\begin{equation}
\rho(\epsilon)=-\frac{1}{\pi}\,\text{Im} \,\Tr \hat{G}(\epsilon)=\rho_0(\epsilon)+\delta\rho_A(\epsilon)
\end{equation} 
where the variation in the density of states can be easily shown to be
\begin{equation}
\delta\rho_A(\epsilon)=-\frac{1}{\pi}\,\text{Im}\,\Tr \partial_\epsilon\,\ln(\hat{I}-\hat{V}_A\,\hat{G}_0(\epsilon))
\end{equation}
The energy variation at zero temperature is
\begin{eqnarray}
\delta E_A^{KS}&=&\int_{-\infty}^{E_F}\,\text{d}\epsilon\,\epsilon\,\delta\rho_A(\epsilon) \nonumber \\
&=&-\frac{1}{\pi}\,\int_{-\infty}^{E_F}\,\text{d}\epsilon\,\text{Im}\,
\Tr \,\ln(\hat{I}-\hat{V}_A\,\hat{G}_0(\epsilon)) \, ,
\end{eqnarray}
where the last step is achieved via integration by parts. We assume now that the perturbation is 
infinitesimal, 
\begin{equation}
\hat{V}_A=\hat{V}_A^1\,\delta\theta+\hat{V}_A^2\,\delta\theta^2
\end{equation}
and expand the energy variation to second order so that
\begin{equation}
\label{eqn:dft1site}
\delta E_A^{KS}=-\frac{1}{\pi}\,\int_{-\infty}^{E_F}\text{d}\epsilon\, \text{Im}\,\Tr\left(\hat{V}_A^2\,\hat{G}_0
+\frac{1}{2}\,\hat{V}_A^1\,\hat{G}_0\,\hat{V}_A^1\,\hat{G}_0\right)\,\delta\theta^2
\end{equation}
If $\hat{V}_A$ refers to a local perturbation then the perturbation matrix elements are
\begin{equation}
\label{eqn:dv1site}
V^{1,2}_A=\begin{pmatrix}V_{AA}^{1,2}&\frac{1}{2}\,V_{AB}^{1,2}\\\frac{1}{2}\,V_{BA}^{1,2}&0\end{pmatrix}
\end{equation}
while the full unperturbed Green's function matrix must be used
\begin{equation}
G_0(\epsilon)=\begin{pmatrix}G_{0,AA}&G_{0,AB}\\G_{0,BA}&G_{0,BB}\end{pmatrix}
\end{equation}
If we approximate the local perturbation by the on-site approximation by discarding the matrix elements $V_{BA},\,V_{AB}$, then the
energy variation can be simplified to
\begin{eqnarray}
\delta E_A^{KS}&=&-\frac{1}{\pi}\,\int_{-\infty}^{E_F}\text{d}\epsilon\, \text{Im}\,\Tr\left(V_{AA}^2\,G_{0,AA}\right. \nonumber \\
&&+\left.\frac{1}{2}\,V_{AA}^1\,G_{0,AA}\,V_{AA}^1\,G_{0,AA}\right)\,\delta\theta^2 \, .
\end{eqnarray}
We suppose now that two different infinitesimal perturbations $A$ and $B$ are applied simultaneously to the system so that
\begin{equation}
\begin{array}{rcl}
\hat{V}_{AB}&=&\hat{V}_A+\hat{V}_B \\ [5pt]
\hat{V}_{A/B}&=&\hat{V}^1_{A/B}\delta\theta+\hat{V}^2_{A/B}\delta\theta^2 \, .
\end{array}
\end{equation}
Then the energy variation up to second order can be recast as 
\begin{eqnarray}
\delta E_{AB}^{KS}&=&\frac{1}{\pi}\,\int_{-\infty}^{E_F}\text{d}\epsilon\,\text{Im}\Tr \ln(\hat{I}-\hat{V}_{AB}\,\hat{G}) \nonumber \\
&=&\delta E^{KS}_A+\delta E^{KS}_B+\delta E_{AB}^{int} \, ,
\end{eqnarray}
where
\begin{eqnarray}
\label{eqn:dft2site}
\delta E_{AB}^{int}&=&-\frac{1}{\pi}\,\int_{-\infty}^{E_F}\text{d}\epsilon\,\text{Im}\Tr (\hat{V}_A^1\,\hat{G}_0\,\hat{V}_B^1\,\hat{G}_0)\,\delta\theta^2 \, .
\end{eqnarray}
If $\hat{V}_{A,B}$ refer to local perturbations, then some extra sorting and book-keeping of the PAO indices must
be performed. We arrange the PAO basis into three subsets $\{A,\,B,\,R\}$, so that the complementary subset to 
$A$ is $\{B,\,R\}$, and to $B$ is $\{A,\,R\}$, and the matrices corresponding to the local perturbation operators are
\begin{eqnarray}
\label{eqn:dv2site}
V^1_A&=&\begin{pmatrix}V_{AA}^1&\frac{1}{2}\,V_{AB}^1& \frac{1}{2}\,V_{AR}^1
\\\frac{1}{2}\,V_{BA}^1&0&0\\\frac{1}{2}\,V_{RA}^1&0&0\end{pmatrix}\\ \nonumber \\
V^1_B&=&\begin{pmatrix}0&\frac{1}{2}\,V_{AB}^1&0\\\frac{1}{2}\,V_{BA}^1&V_{BB}^1&\frac{1}{2}\,V_{BR}^1\\
0&\frac{1}{2}\,V_{RB}^1&0\end{pmatrix} \, .
\end{eqnarray}
Within the on-site approximation for the local perturbation the matrix elements other than 
$V_{AA},\,V_{BB}$ are neglected, thus the energy variation can be simplified to
\begin{eqnarray}
\delta E_{AB}^{int}&=&-\frac{1}{\pi}\,\int_{-\infty}^{E_F}\text{d}\epsilon\,\text{Im}\Tr (V_{AA}^1\,G_{0,AB}\,V_{BB}^1\,G_{0,BA}) \delta\theta^2 \, . \nonumber \\
\end{eqnarray}
If the local perturbation consists of a spin rotation of the XC potential $H_{XC}={\bf V}_{XC}\cdot{\boldsymbol \tau}$ 
around an axis ${\bf u}$, the perturbed potential matrices are, 
\begin{equation}
\label{eqn:dvspin}
\begin{array}{rcl}
\delta V^{{\bf u},1}&=&\,\displaystyle \frac{i}{2}\,[\,H_{XC},T^{\bf u}\,] \\ [8pt]
\delta V^{{\bf u},2}&=& \displaystyle \frac{1}{8}\,[\,[\,T^{\bf u},H_{XC}\,],T^{\bf u}\,] \, .
\end{array}
\end{equation}
The DFT2S mapping finishes by comparing Eq. (\ref{eqn:singlesite}) to Eq. (\ref{eqn:dft1site}) for single spin rotations at site $A$,
and Eq.(\ref{eqn:doublesite}) to Eq. (\ref{eqn:dft2site}) for double spin rotations at sites $A$, $B$, whereby all magnetic constants
can be extracted.

\subsection{Implementation of the LKAG method}
We describe in this section the specific decisions and steps that must be taken to implement the LKAG torque method. These are

{\it Step 1}. Perform a DFT simulation where a given quantization axis ${\bf o^1}$ has been selected so that 
${\bf V}_{XC}\parallel {\bf o^1}$. We assume that ${\bf o^1}$ lies along one of the three Cartesian axes 
defined for the classical Hamiltonian. Obtain the KS Hamiltonian $H^1$. \newline

{\it Step 2}. Extract ${\bf V}_{XC}^1$ from  $H^1$.
Rotate ${\bf V}_{XC}^1$ to align it with each of the two Cartesian axes perpendicular to ${\bf o^1}$, that we call
${\bf o^2}$ and ${\bf o^3}$. Use the new exchange fields ${\bf V}_{XC}^{2/3}$ to determine the KS Hamiltonians
\begin{equation}
H^{2/3}=H^{TRS}+\frac{1}{2}\,{\bf V}_{XC}^{2/3}\cdot {\boldsymbol \tau} \, .
\end{equation}
We are then furnished with a reference Hamiltonian quantized along each of the three Cartesian axes, that complies
with the requisite that the spin-orbit part of the Hamiltonian should not be rotated. These three Hamiltonians are needed
to achieve the mapping and find all the matrix elements of the magnetic tensors.
Accordingly, the next steps in the mapping must be performed for each of the three quantization
axes ${\bf o}^{1/2/3}$, hence the three KS Hamiltonians.  

{\it Step 3}. Choose one of the three quantization axis (we call it ${\bf o}$ without the superindex to simplify 
the notation) and define two perpendicular axes ${\bf v}$ and 
${\bf w}={\bf o}\times {\bf v}$. Perform an infinitesimal rotation for each magnetic site $A$ in the unit cell, 
around each of the two axes ${\bf v}$ and ${\bf w}$, thus obtaining $\delta V^{{\bf v}, {\bf w}}_A$ by
Eq. (\ref{eqn:dv1site}) and Eq. (\ref{eqn:dvspin}). The energy variation is obtained using Eq. (\ref{eqn:dft1site}).

{\it Step 4}. For a given magnetic site in the unit cell $A$, we identify the list of neighboring magnetic
sites $B$ such that the exchange tensor $J_{AB}$ is non-negligible. For each of these sites $B$, we perform double-site 
rotations around the axes ${\bf v}_A$, ${\bf w}_B$, that
gives a total of four rotations for each quantization axis. The energy variation for each is obtained from 
Eq. (\ref{eqn:dft2site})
and Eq. (\ref{eqn:dv2site}) and Eq. (\ref{eqn:dvspin}). The number of equations (12) is larger than the number of 
unknown  quantities (9), which enables us to perform consistency checks. 

We comment now on an additional issue related to the identification of the electron degrees of freedom that 
can be considered as localized because they have an energy gap in the charge and longitudinal spin sectors,
as we have discussed in the Introduction.
We have found that the quality of the determined exchange and anisotropy constants depends dramatically
on the choice of PAO orbitals that are considered localized at sites $A$, $B$ and will hence
be rotated via Eq. (\ref{eqn:dvspin}). So much so that the parameters may change by large factors and even by
orders of magnitude in some cases if PAO orbitals representing non-localized electrons are included, as will 
be discussed further below.
 
\section{Tests and benchmarks on diverse nanostructures}
\label{section:Tests}
We have implemented the DFT2S mapping described in the previous sections as a post-processing package 
\cite{Grogu} of the PAO-based DFT program SIESTA\cite{siestapaper,siesta20}. We devote this section to disclose and
comment on some of the tests that we have carried out to ensure that the proposed approach is satisfactory.

The first two subsections detail consistency tests that we have performed on two simple systems. These are 
an isolated platinum dimer and an isolated chromium trimer. We mention that 
the apparent simplicity of those two systems has allowed us to carry out stringent tests, that might be overlooked 
or hidden in more complex systems.
We have tested for these systems the performance of the local and on-site projections, as well as the impact of
rotating either the full KS Hamiltonian or only the exchange field vector ${\bf V}_{XC}$ on the DFT2S mapping.
We have used as guiding principles the dimer sum rule established in Eq.  (\ref{eqn:dimersumrule}) and other
simple symmetry considerations. As an example, we expect that a rotation about the quantization axis should lead
to a zero energy variation.
We have picked the Hamiltonian of the above two systems that correspond to a $z$-quantization axis, 
and have rotated ${\bf V}_{XC}$ to have it aligned along the $x$- and the $y$-axes. 
We have computed the spin vector again and have found that the modulus is still different. We have then 
decided to compute the available constants for the magnetic multiplet corresponding to the $z$-quantization 
axis.

We have realized while implementing and testing our code that the PAO radii have a strong impact on the numerical
values of the exchange parameters. Interestingly enough, we have found that the default cutoff radii provided by 
the program SIESTA tend to deliver reasonable estimates of the exchange parameters. However, those cutoff radii are
variational parameters that can be improved by minimizing the total KS energy, which tends to deliver radii longer
than SIESTA's defaults. We have found that the exchange parameters obtained whenever using the optimized radii 
differ easily by factors of 4-5 or even by orders of magnitude, if all PAOs are included in the atomic rotations.
This issue is circumvented if only the PAOs related to localized degrees of freedom are rotated, which in the cases 
below mean those describing $d-$orbitals. We have tested that in this case, the exchange constants bear small modifications when the PAO radii are changed. The variational optimization of the PAO radii can then be used 
as a guiding principle to improve the quality of the exchange constants.

We have carried  two detailed benchmarks against the more established KKR framework\cite{Udvardi_relativistic_Jij},
where we have analyzed two nanostructures deposited on metallic surfaces. We have found that the two 
approaches agree very well for the exchange tensors. The DM vectors show slightly larger discrepancies, that
we have been able to trace down to differences in the electronic structure, that can be identified while
analyzing the Projected Densities of States (PDOS) of those systems.

The accuracy parameters common to all the simulations discussed in this section are as follows. We have employed 
a double-$\zeta$ polarized basis set whose first-$\zeta$ PAOs had long radii of about 8 Bohr. We have optimized 
the pseudopotentials by a fitting procedure to the electronic band structure, lattice constants and
magnetization of the plane-wave code VASP, that we have described elsewhere\cite{Rivero2015}. 
We have used a Local Spin Density Approximation XC functional \cite{PZ-LSDA} whenever we have benchmarked our results 
against the KKR method. We have used the more adequate PBE functional otherwise\cite{PBE}. We have used a stringent 
set of accuracy parameters due to the strong sensibility  of anisotropy-related magnitudes whenever the 
spin-orbit interaction has been included. As an example, we have used
real-space mesh grids equivalent to a plane-wave energy cutoff of 1000-1200 Ry, depending on the case.
The electronic temperature for the smearing occupation function has been set typically as low as 0.1 K to
identify and discern small degeneracy liftings due to the spin-orbit interaction. Similarly, the tolerance
on the  error of the density matrix and the Hamiltonian matrix elements has been set as low as 
$10^{-6}$ and $10^{-5}$ eV, respectively. We have been able to converge a collinear FM solution for all the 
systems described in this work, and we have taken it as the local/global energy minimum state upon which we 
perform  the DFT2S mapping. 
We have used simulation boxes with a lateral length as large as 20 \AA\ to avoid finite-size effects for the platinum 
dimer and the chromium trimer. 

\begin{table}[h]
\caption{\label{tablesumrule} Relative error committed in the sum rule in Eq. (\ref{eqn:dimersumrule}) 
for a platinum dimer, where both the local and the onsite projections are tested.
The sum rule is fulfilled up to numerical accuracy for the local projection when all PAOs are taken 
into account. We also find that the error increases to 3.8\% when the PAO basis is truncated to $s$ and $d$ orbitals. 
In sharp contrast, the relative error in the sum rule grows to about $14\%$ when the on-site projection is used.}
\begin{tabular*}{0.5\textwidth}{@{\extracolsep{\fill}}ccc}
 \hline
 & Local projection &  On-site projection   \\
 \hline
 All PAOs    & 0   &14.2  \\
  $sd$ PAOs  & 3.8 & 13.9  \\
 \hline
\end{tabular*}
\end{table}

\subsection{Isolated Pt dimer}
\label{IsolatedPtdimer}

We have placed a platinum dimer at the DFT equilibrium distance along the $z$-axis, as determined by the code 
SIESTA. We have been able to converge DFT simulations for the dimer,  where the quantization axis 
was oriented along each of the three coordinate axes. We have computed total 
energy differences that are consistent with previous published results \cite{FF07}. 
The total angular momenta per atom are ${\bf J}_A=(1.8, 0, 0)$, $(0,1.8,0)$ and $(0, 0 , 4)$ $\mu_\mathrm{B}$ for 
the $x$, $y$ and $z$ quantization axes, respectively. Similarly, the spin moments are ${\bf S}_A = (0.98,0,0)$, $(0,0.98,0)$, and $(0,0,3.06)$ $\mu_\mathrm{B}$.
These values indicate that the three states belong to different magnetic multiplets, so that the DFT2S mapping leads
to three different Heisenberg Hamiltonians.

We have then analyzed whether the sum rule in Eq. (\ref{eqn:dimersumrule}) is verified for a 
platinum dimer. We have found that our definition of a localized operator fulfills that sum rule, while the 
conventional on-site projection fails to fulfill it by about 14 \% as shown in Table \ref{tablesumrule}.

We have then performed rotations around the quantization axis to check that the computed energy 
variation is zero, as it should by symmetry. We have found that rotating the full Hamiltonian leads to non-zero energy 
values as shown in Table \ref{tabledimerrotation}. In contrast, the energy difference is reduced to zero up 
to numerical accuracy if only the exchange field is rotated. By numerical accuracy we mean here energy differences 
within the range $0.01\,$ meV. The second column in Table \ref{tabledimerrotation} shows that 
rotating ${\bf V}_{XC}$ or the full Hamiltonian produces qualitative differences on the computed values of the 
exchange constants that 
must not be overlooked.

\begin{table}[h]
\caption{\label{tabledimerrotation} Impact of rotating the full Hamiltonian $H^{KS}$, or only 
the exchange 
field ${\bf V}_{XC}$ for a platinum dimer. We denote the two dimer atoms by $A$ and $B$. The first column shows 
the second-order single-site rotation energy $E^2_A\left(\{{\bf u}_z\},{\bf u}_z\right)$, that should be identically zero 
by symmetry. The second column displays the  exchange constants $J^{yy}_{AB}=J^{xx}_{AB}$ for the same dimer. All quantities are displayed using the local / on-site projection and measured in meV.}
\begin{tabular*}{0.5\textwidth}{@{\extracolsep{\fill}}ccc}
 \hline
 &  $E^2_A\left(\{{\bf u}_z\},{\bf u}_z\right)$ & $J^{yy}_{AB}=J^{xx}_{AB}$  \\
 \hline
Rotating $H$ & -19 / -22 & -134 / -138  \\
Rotating ${\bf V}_{XC}$  & 0 / 0& -65 / -70  \\
 \hline
\end{tabular*}
\end{table}

\begin{table}[h]
\small
\caption{\label{Table:table_Cr3_dorbitals}
$J^{yy}_{12}$ for the isolated Cr trimer analyzed in section \ref{subsection:isolatedtrimer} for 
different input energy shifts in mRyd - a variable that controls the cutoff radii $R^1_{s,p,d}$ of the first-zeta PAO. As a rule of thumb, smaller 
energy shifts lead to larger PAO radii. Larger PAO radii tend to be better from a variational point of view.
All exchange energies are given in meV.} 
\begin{tabular*}{0.48\textwidth}{@{\extracolsep{\fill}}cccc}
 \hline
Energy shift & $R^1_{s/d}$ (Bohr) & $J^{yy}_{12}$ (all PAOs) & $J^{yy}_{12}$ ($d$ PAOs)\\
 \hline
 0.1  & 12.4 / 9.9 & 625 & 154 \\
 1    & 10.2 / 7.5 & 516 & 154 \\
 10   &  7.7 / 5.2  & 260 & 130 \\
 20   &  7.0 / 4.6  & 143 & 109 \\
 \hline
\end{tabular*}
\end{table}

\begin{table}[h]
\small
  \caption{\label{Table:table_Cr3_dorbitals_splitnorm}
 $J^{yy}_{12}$ for the isolated Cr trimer studied in section \ref{subsection:isolatedtrimer} for the same first-zeta PAO radii 
 ($R^1_s=10.2$ Bohr, \ $R^1_d=7.5$ Bohr)  and different split norms, a variable that controls the cutoff 
 radii $R^2_{s,d}$ of  the second-zeta PAO. }
 \begin{tabular*}{0.48\textwidth}{@{\extracolsep{\fill}}cccc}
 \hline
 Split Norm & $R^2_{s,d}$ (Bohr) & $J^{yy}_{12}$ (all PAOs) & $J^{yy}_{12}$ ($d$ PAOs)\\
    \hline
 0.05 &  8.8, 4.4       & 596 & 163 \\
 0.10 &  7.8, 3.6       & 555 & 156 \\
 0.15 &  7.3, 3.1       & 507 & 155 \\
 0.20 &  6.9, 2.7       & 494 & 154 \\
 0.25 &  6.6, 2.5      & 494 & 154 \\
    \hline
\end{tabular*}
\end{table}

\subsection{Isolated equilateral chromium trimer}
\label{subsection:isolatedtrimer}
\begin{figure}[ht]
    \centering
    \includegraphics[width=\columnwidth]{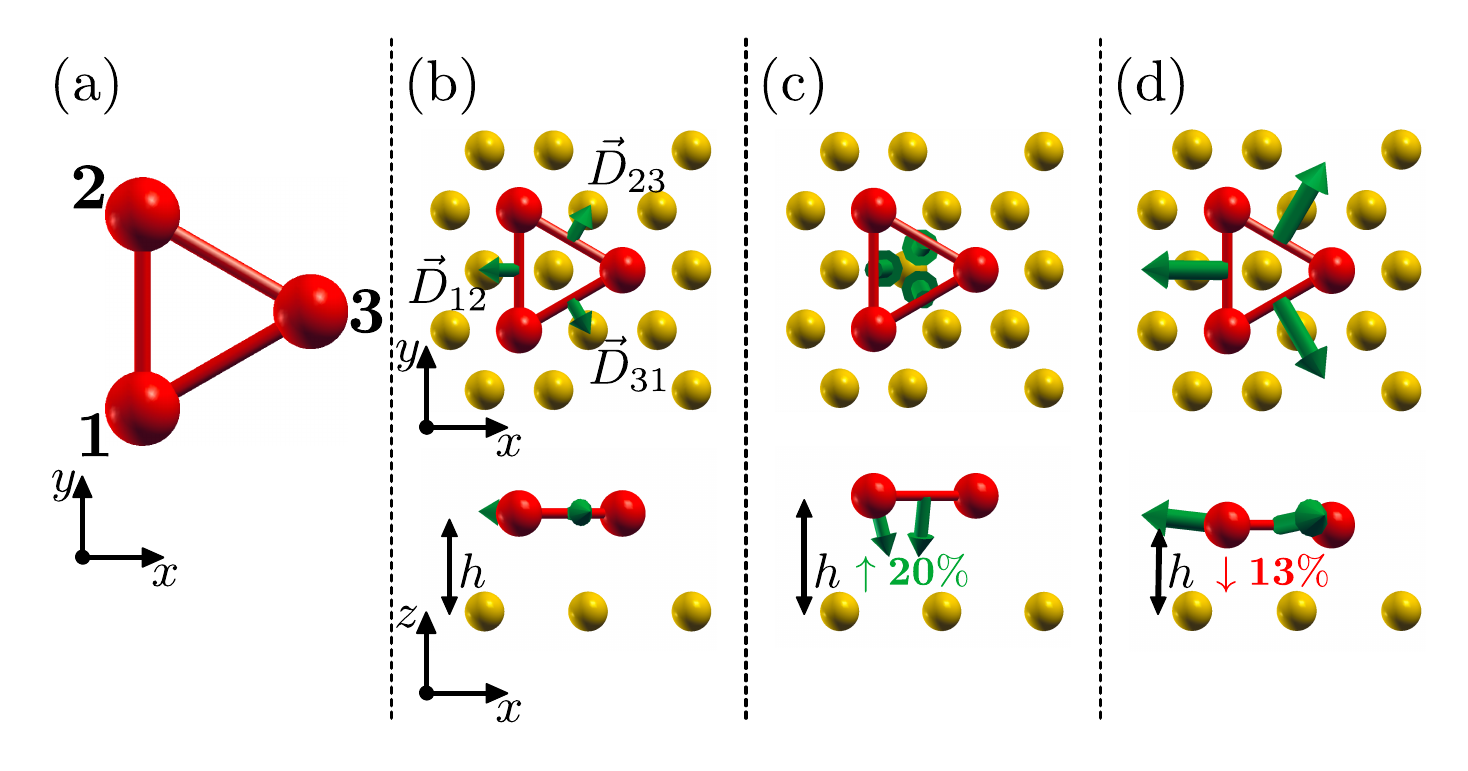}
    \caption{Schematic drawings of the simulated chromium trimers. (a) the isolated trimer, where the convention 
    for the coordinate axes is indicated; (b), (c) and (d) top and side views of the Cr trimer/Au (111) 
    heterostructure for three different heights  $h=2.36$\,\AA\, $h=2.83$\,\AA\; and $h=2.06$\,\AA.  
    Red- and gold-colored spheres denote the Cr and the surface Au atoms, respectively. The figure also 
    shows the DM vectors in green arrows. The DM magnitudes are rescaled among the three figures to allow 
    for a better visualization of the vectors. }
    \label{fig:fig1}
\end{figure}

We analyze now the exchange tensors between the different atomic pairs of the chromium trimer shown 
in Fig. \ref{fig:fig1} (a). 
We have arranged the chromium atoms to form an equilateral triangle lying in the $xy$-plane. We have found
local energy minima for the three quantization axes ${\bf x},\, {\bf y},\, {\bf z}$. However, we have checked that they 
have different atomic angular moments, so that they belong to different multiplets. We have picked the lowest-energy 
state among the three, that corresponds to the spins aligned along the $z$-axis, at a distance of 2.885 \AA \ between the Cr atoms. 
The atomic spin moment at this energy minimum is $5.33 \, \mu_{\rm B}$ and the orbital contribution is 
$0.166 \, \mu_{\rm B}$, giving $m_J=5.5 \, \mu_{\rm B}$ per atom, and a total spin moment of 
$16 \, \mu_{\rm B}$ for the trimer. The DFT2S mapping then allows us to infer only 
$J^{xx},\,J^{yy}$, $J^{xy,S}$, $D^z$, $K^{xx}-K^{zz}=K^{yy}-K^{zz}$, $K^{xz}$ and $K^{yz}$.

We have verified for the trimer that (a) the DM sum rule in Eq. (\ref{eqn:DMsumrule}) is 
fulfilled for the 
$D^z$ component; (b) that $D^z_{ij}$ is the same for the three atom pairs; (c)
that the intra-atomic anisotropy matrix elements $K_i^{xx}-K_i^{zz}$ of the
the atoms are related by $2\,\pi/3$ rotations; (d) that $J_{ij}^{S,\alpha\beta}$ 
for the three atom pairs are related by $2\,\pi/3$ rotations around the $z$-axis. 

We have first analyzed the dependence of the exchange constant $J^{yy}_{12}$ on the PAO radii,
where we have used the local projection proposed in this article. Table \ref{Table:table_Cr3_dorbitals} shows 
that $J^{yy}_{12}$ changes by a factor of 4 to 5 when modifying the radii of the first-$\zeta$ PAO if all the PAOs 
are included in the infinitesimal perturbation. In contrast, the change is much smaller if only the magnetic $d$ 
orbitals are taken into account. Table \ref{Table:table_Cr3_dorbitals_splitnorm} shows that the radii of the 
second-$\zeta$ on $J$ has in contrast a very small impact on $J_{12}^{yy}$.

The calculated exchange and anisotropy parameters for the atom pair $(1,2)$ are shown in Table~\ref{table_Cr_vacuum}.
The diagonal exchange interactions between the Cr atoms are antiferromagnetic as expected, 
leading to the well-known geometric frustration of the chromium trimer.  
The intra-atomic anisotropy, $K_i^{xx}-K_i^{zz} \simeq K_i^{yy}-K_i^{zz}$ is sizable and 
positive ($\simeq 14$ meV).

\begin{table}[h]
\small
 \caption{\label{table_Cr_vacuum}
Exchange and anisotropy constants of the atom pair (1,2) of an isolated equilateral Cr trimer. 
The table shows the results obtained using both the local projection and the on-site projection. 
All quantities are measured in meV.}
 \begin{tabular*}{0.48\textwidth}{@{\extracolsep{\fill}}lccccc}
    \hline
    &                         $J_{12}^{xx}$  &  $J_{12}^{yy}$  &  $J_{12}^{xy,S}$   &$D_{12}^{z}$     &$K_1^{xx}-K_1^{zz}$   \\
    \hline
 Local Projection    & 155      &    155        &    0             &    -14         &       14                \\
 On-site Projection  & 134      &    134        &    0             &     -14        &       16                \\
    \hline
 \end{tabular*}
\end{table}

\begin{table}[ht]
\small
  \caption{\label{table_Cr3}
$J^H$, $D^z_{12}$ and $D^x_{12}$ for the equilateral Cr trimer deposited on top of a Au (111) surface, 
for the three different heights shown in Figure (\ref{fig:fig1}). The table shows the results calculated with both 
the local projection and the on-site projection, using the format LP / OP. All data are given in units of meV.}

  \begin{tabular*}{0.48\textwidth}{@{\extracolsep{\fill}}cccc}
    \hline
    $h$\,(\AA) &  $J^H$    &  $D^z_{12}$  &  $D^{y}_{12}$\\
    \hline
     2.83      & 177 / 157 & -3  / -2.5   &  0.8 / 0.6 \\
     2.36      & 159 / 143 & 0.1 / 0.3    & -1.9 / -2 \\
     2.06      & 145 / 131 & 1.2 / 0.8    & -7.2 / -7 \\
    \hline
  \end{tabular*}
  \label{tableJD_trimer}
\end{table}

\subsection{Cr trimer deposited on a Au(111) surface}
\label{subsection:supportedtrimer}
We have simulated a heterostructure consisting of four atomic layers of gold atoms stacked along the (111) direction. 
The simulation box includes $5\times 5$ gold atoms in each layer. Periodic boundary conditions are applied in each direction,
where we have added enough vacuum space along the $z$-axis to detach the heterostructure from its periodic images. We have 
used an LDA functional, and have set the lattice constant equal to $a_{\rm 2D}=2.885$\,\AA. The $k$-grid in the DFT simulation
has been restricted to just the $\Gamma$-point. We have then placed the chromium trimer onto the top gold layer at the 
hollow sites, as indicated in Fig. (\ref{fig:fig1}) (b). Here, the chromium intra-atomic distance has been set to match with gold lattice 
constant (we note that the lattice mismatch is actually very small). We have however placed the trimer at three different heights $h$ to check the impact of vertical strain of the 
exchange and anisotropy constants. These are (a) the ideal fcc(111) interlayer distance, 
$h=\frac{\sqrt{6}}{3} a_{\rm 2D}=2.36$\,\AA\,, (b) $h=2.83$\,\AA\ that corresponds to an outward relaxation of  $20\,\%$,
and (c) $h=2.06$\,\AA\, that corresponds to an inwards relaxation of $13\,\%$ and it is the optimum height obtained by relaxing the forces towards the gold substrate.
We have simulated a ferromagnetic 
spin configuration with spins pointing along the $z-$axis. We have then rotated ${\bf V}_{XC}$ and verified that the spin moments
do not change. Hence, spin rotations do not change the magnetic multiplet in this case. The single-site sum rule allows
us to extract only the off-diagonal matrix elements $K_i^{xz}$ and $K_i^{yz}$, but not $K_i^{xy}$. The second-order single-site 
variations allow us to determine the diagonal matrix elements $K_i^{xx}$ and $K_i^{yy}$. The double-site second order energy variation
allows us to gain access to all the exchange tensor matrix elements $J_i^{\alpha\beta}$. 
The missing matrix element $K_i^{xy}$ can be found by performing an additional rotation about a unit vector lying in the XY plane that
is not collinear neither to ${\bf x}$ nor to ${\bf y}$. Alternatively, we can exploit the symmetries among the three chromium atoms,
that allow us to relate $K_{2}^{xx},\ K_2^{yy}$ to $K_1^{xy}$ and so on, by performing $2\,\pi/3$ rotations around the $z$-axis on each 
$\tilde{K}_i$ matrix.

We discuss first our results for the ideal fcc(111) interlayer spacing, $h=2.36$\,\AA. We have found here that the spin moment of 
each Cr atom is $4.709\, \mu_{\rm B}$ and that the orbital moment is very small, $0.003 \, \mu_{\rm B}$. The reduction of the spin 
moment with respect to the isolated Cr trimer is attributed to the hybridization between the Cr and Au $d$-orbitals.  

We have found that the 
three diagonal components of the exchange tensor are equal to each other, within a margin of $\pm 0.3$ meV. 
Furthermore, they are equal for the three chromium atoms within the same margin, so we denote them simply $J^H$. 
The symmetric exchange matrix elements have values smaller than 0.3 meV. 
Table \ref{table_Cr3} shows $J^H$, together with the DM vector components $D^z_{12}$ and $D^y_{12}$ of the atom pair
(1,2). 
We have checked that the DM vectors for the pairs (2,3) and (3,1) obtained via rotations of ${\bf D}_{12}$ 
by $120^\circ$ and  $240^\circ$ around the  $z$-axis, respectively, agree with the direct calculations. 
The Table shows that the estimates provided by the on-site projection differ from those obtained by the local projection 
by 10 \% for $J^H$ and 20-25 \% for the DM vectors. 

The intra-atomic anisotropy matrix for atom 1 obtained using the on-site projection is
\begin{eqnarray}
 \tilde{K}_1&=&\begin{pmatrix}
   -0.06       &  K_1^{xy}  & -0.8\\
   K_1^{xy}    &  -0.10     &  1.6\\
   -0.8        &   1.6      &  0
\end{pmatrix} \mathrm{meV}
\end{eqnarray}
We have determined that $K_1^{xy}=-0.0346$ meV by rotating $\tilde{K}_2$ by an angle $2\,\pi/3$ around the $z$-axis.

Table \ref{table_Cr3} shows how decreasing/increasing the trimer height reduces/increases $J^H$,  which can again be ascribed to
the change in the localization character of the eigenstates with dominant Cr-$d$ orbital contribution. We find an interesting and intriguing rotation of the 
DM vectors when the height is changed as shown in Fig. \ref{fig:fig1}:  for the largest height, 
the DM vectors point almost perpendicular to the plane of the triangle as in the isolated Cr trimer; the DM vectors rotate gradually 
towards the surface plane and point outwards from the triangle center as the height is reduced

We have chosen the trimer height $h=2.36$\,\AA\ to be able to benchmark our results against the established embedded cluster KKR Green's function 
technique\cite{Lazarovits2002} and the relativistic torque method\cite{Udvardi_relativistic_Jij}. 
We have used here a cut-off of $\ell_{\rm max}=3$ for the partial waves in multiple scattering, an LDA functional and the Atomic Sphere
Approximation for the potentials. We have found here that all diagonal components are similar among them and for all three pairs, with small differences of about 0.3 meV, in agreement with the results of our approach. We
have calculated that $J^H=143$\,meV, that agrees very well with Table \ref{table_Cr3}. In contrast, we have found a larger disagreement
for the DM vectors, since KKR finds $D^z_{12}=-0.95$\,meV and $D^{y}_{12}=1.2$\,meV. 
This discrepancy puts forward the large sensitivity of the DM vectors against the details of electronic structure. The observed 
differences might in particular be due to the fact that the  KKR calculations avoided using a supercell approach by employing a
 semi-infinite Au(111) substrate and setting the Fermi level from truly bulk calculations.

\begin{figure}[ht]
    \centering
    \includegraphics[width=\columnwidth]{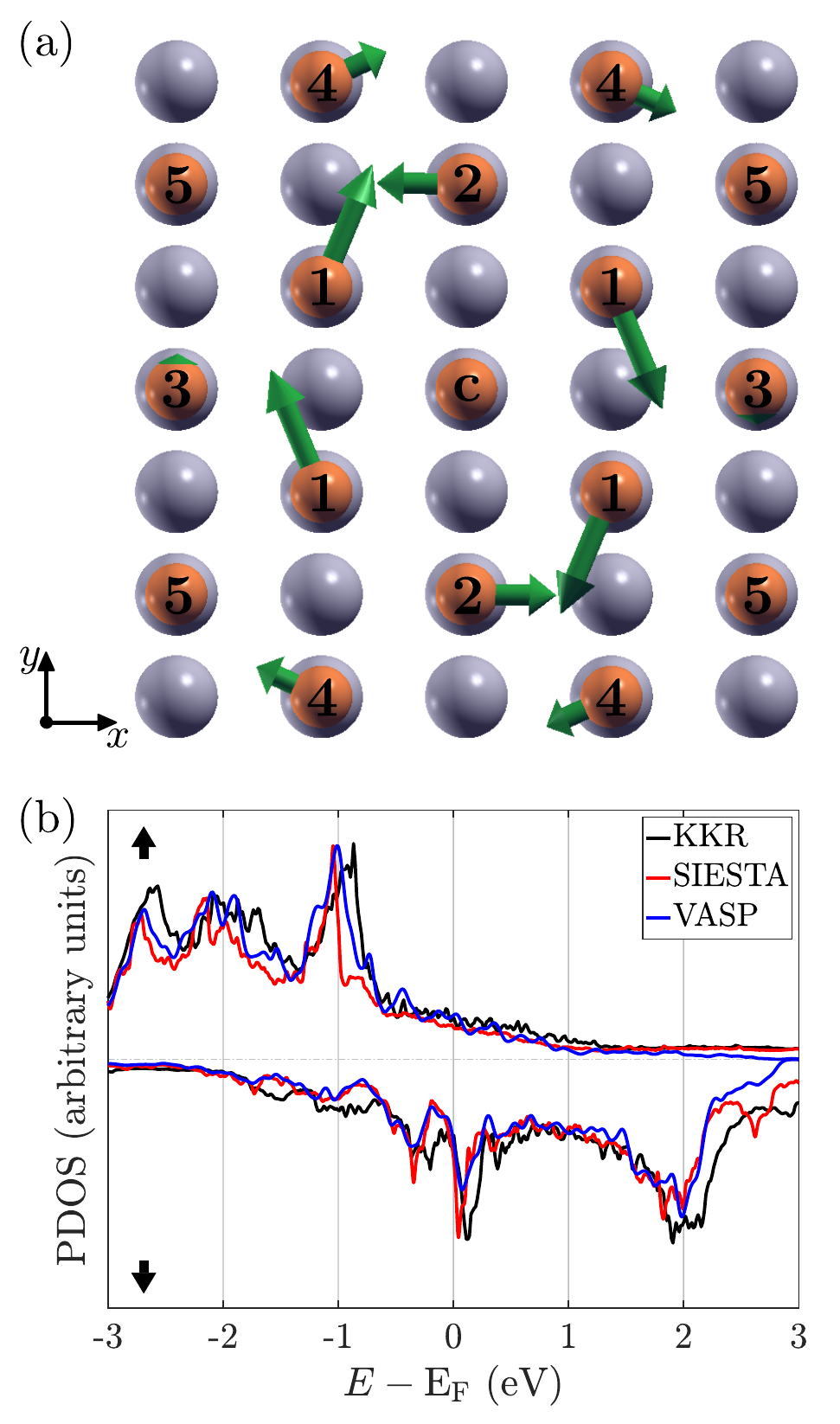}
    \caption{(a) Top view of the simulated Mn monolayer deposited onto a W(110) slab. 
    Orange/grey spheres correspond to Mn/W atoms. Note that the W atoms below the Mn atoms occupy sites in 
    the subsurface W layer. The DM vectors referred to the central Mn atom (labeled by the letter c) 
    are depicted in green color. All the DM vectors lie within the surface plane. Their moduli from the first to 
    the fifth nearest neighbor are 3.7, 2.5, 1, 2 and 0.7 meV. (b) PDOS projected onto the Mn atom, obtained from
    a collinear spin calculation. The top/bottom panels refer to spin-up/-down projections, respectively. Black, red and blue
    lines refer to results from the codes KKR ($\ell_{\rm max}=3$), SIESTA and VASP.}
    \label{fig:Mnw}
\end{figure}

\begin{table}[h]
\small
  \caption{\label{table_MnW}
Calculated isotropic Heisenberg interactions $J^H_{CN}$ for the five nearest neighbor Mn-Mn pairs in the 
Mn/W(110) heterostructure. All data are given in units of meV. }
  \begin{tabular*}{0.48\textwidth}{@{\extracolsep{\fill}}lllllc}
    \hline
    &$J^{H}_{\text{C1}}$&$J^{H}_{\text{C2}}$&$J^{H}_{\text{C3}}$&$J^{H}_{\text{C4}}$&$J^{H}_{\text{C5}}$\\
    \hline
 Local projection& 83.4 & -38.3&-11.1&22.4&-17.1  \\
 On-site projection& 73.0 & -46.0&-13.0&22.9&-12.9 \\
 KKR& 88.0 & -60.4&-4.2&28.1&-11.8\\
    \hline
  \end{tabular*}
\end{table}

\subsection{Mn monolayer on W(110)}
We analyze now a heterostructure where large DM vectors have been found to generate a spin spiral ground 
state \cite{UDVARDI2008402,PhysRevB.91.064402}. The heterostructure is depicted schematically in Fig. \ref{fig:Mnw} (a), 
and consists of a W(110) surface where a Mn monolayer has been deposited. We assume that the Mn monolayer grows epitaxially 
on the bcc(110) surface of the W substrate with a lattice constant along the $y$-axis of $a=3.165$\,\AA. We have modeled 
the system by a supercell containing ten atomic layers of W and the Mn monolayer on top. We have chosen an LDA exchange-
correlation functional. We have obtained a Mn spin moment of $3.52 \,\mu_{\rm B}$ and a small orbital angular moment 
of $0.007\,\mu_{\rm B}$. These values are in good agreement with those obtained from the KKR approach ($3.43 \,\mu_{\rm B})$. 

We have found that the Heisenberg exchange $J^H_{c1}$ gives the largest contribution  by a factor of two orders of magnitude
so we only discuss this exchange constant.
Table~\ref{table_MnW} shows the calculated $J^H$ constants for the five nearest Mn-Mn neighbors, for the local projection, the on-site projection and for the KKR approach. We find that the three set of data agree reasonably
well with each other. We find that  $J^H$ shows an oscillatory pattern with nearest-neighbors antiferromagnetic couplings.

The symmetric exchange tensor for nearest neighbor pairs is
\begin{eqnarray}
\label{bilayertensors}
 J_{c1}^{S}&=&\begin{pmatrix}
    0.6   &  -0.3  &  0\\
   -0.3   &  -0.4  & 0\\
    0      &   0     & -0.2
\end{pmatrix} \mathrm{meV}
\end{eqnarray}
while the diagonal intra-atomic anisotropy matrix elements are $K_c^{xx}=-1.4$ meV and $K_c^{yy}=  0.2$ meV. 
This result is consistent with earlier KKR calculations \cite{PhysRevB.91.064402} where the easy and hard 
axes were found to lie along the $x$ (1$\overline{1}0$) and $z$ (110) directions, respectively.

The DM vectors are drawn in Fig. \ref{fig:Mnw} (a). We have checked that they satisfy the sum rule in Eq. (\ref{eqn:DMsumrule}). The figure shows that they lie within the surface plane, and that show
a vortex-like pattern. A very similar vortex pattern has also been obtained previously using the KKR approach 
\cite{PhysRevB.91.064402}, with small discrepancies. The source of those discrepancies lies in the fact that the 
DM vectors depend sensitively on the details of the electronic structure. We plot in Fig. \ref{fig:Mnw} (b) the density
of states projected onto the Mn atoms (PDOS) as computed from SIESTA, KKR and VASP\cite{vasp}. We have found that
the PDOS obtained from SIESTA and VASP agree remarkably well with each other, while KKR has peaks that show slight
upwards energy shifts. These can be attributed to the semi-infinite slab geometry used here, in contrast to the supercell
approach used in the SIESTA and VASP calculations. 
This differences can also explain why the values obtained for $J^H$, albeit qualitatively similar, have discrepancies
of about 10 \%, but that can reach 30 \% for the worst cases.

\begin{figure}[ht]
\centering
\includegraphics[width=\columnwidth]{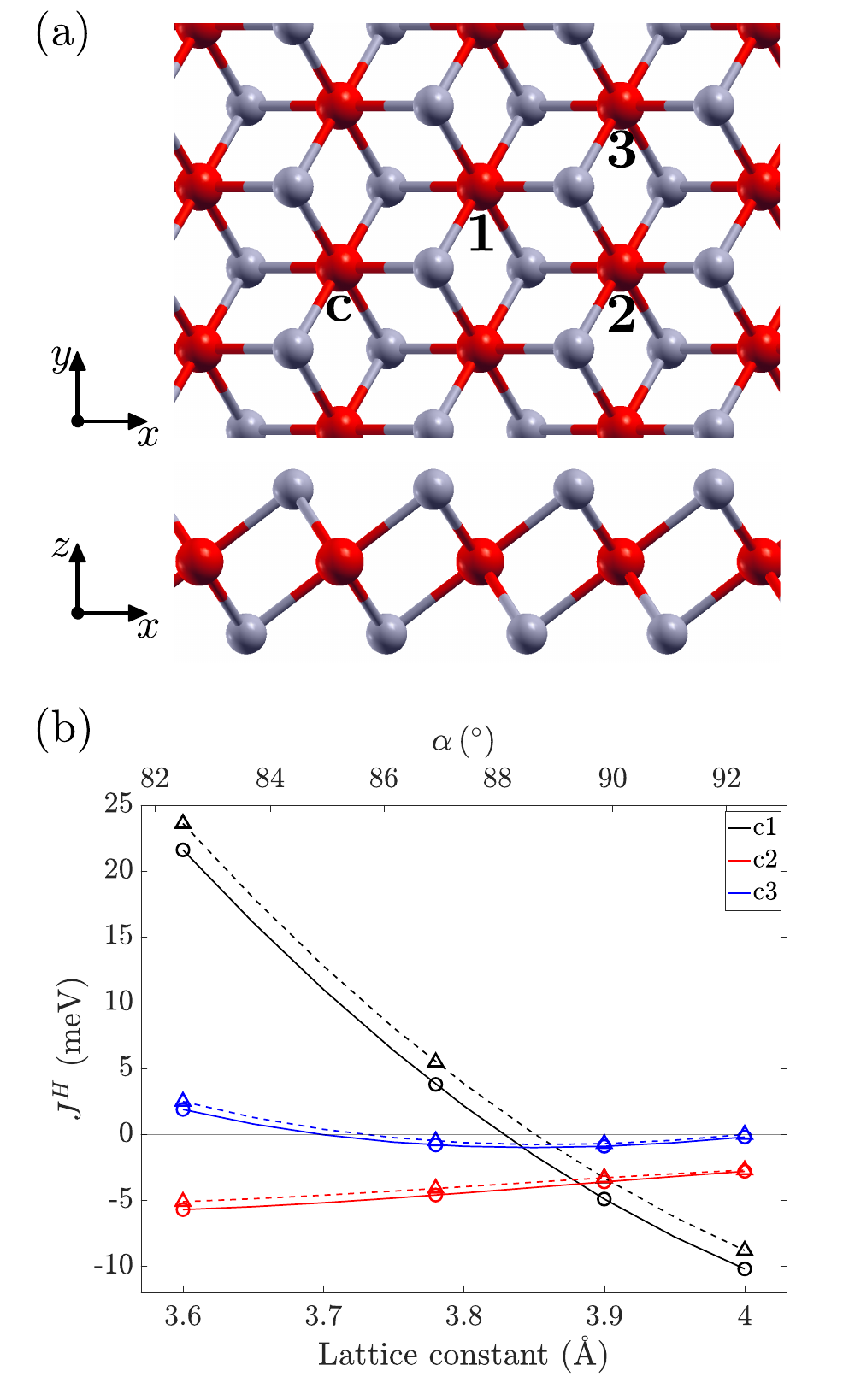}
\caption{(a) Top and side views of the T-CrTe$_2$ crystal structure. Red (grey) colors correspond to Cr (Te) 
atoms. First, second and third neighbors relative to the central c atom are marked as 1, 2 and 3, respectively. 
(b) Isotropic exchange coupling $J_H$ for first, second and third neighbors as a function of the lattice 
constant are plotted with black, red and blue solid/dotted lines.  corresponding to the local/on-site 
projection, respectively. The angle $\alpha$ subtended by the Cr-Te-Cr bonds is plotted at the top $x$-axis.  }
\label{FigCrTe2}
\end{figure}

\section{Impact of strain on the magnetic ground state of T-CrTe$_2$}
\label{section:Crte2}
In this section we investigated the impact of biaxial strain on the exchange parameters of the 1T phase of the magnetic van der
Waals material CrTe$_2$, whose lattice structure is depicted schematically in Fig. \ref{FigCrTe2} (a). The material 
has attracted strong interest \cite{doi:10.1021/acs.jpcc.0c04913,Freitas_2015,PhysRevB.92.214419,PhysRevB.106.L081401}
because the Kanamori-Goodenough-Anderson rules 
\cite{PhysRev.100.564,KANAMORI195987,PhysRev.115.2} indicate that its antiferromagnetic ground state 
could be switched to a ferromagnetic state by the application of biaxial strain that changes the angle subtended by each Cr-Te-Cr bond.
The mechanism is rationalized as follows: for short Cr-Cr distances the direct AFM exchange coupling dominates, however, when 
the distance between Cr atoms increases and Cr-Te-Cr angle approaches 90$^{\circ}$ the FM Cr-Cr 
super-exchange interaction mediated by the intermediate Te atom increases and eventually becomes 
dominant \cite{CHEN201560,Freitas_2015}. 

We have performed DFT simulations of a 1T-CrTe$_2$ monolayer using SIESTA and the PBE GGA functional. We found a theoretical
lattice constant $a=3.78$\, \AA, that agrees fairly well the experimental estimates \cite{Zhang2021,Meng2021}.
We also found that the distance between the Te and Cr planes is $1.68$ \AA. We have used a simple unit cell containing 
one chromium and two tellurium atoms. Therefore we have not analyzed charge density wave distortions, that require a larger unit 
cell\cite{doi:10.1021/acs.jpcc.0c04913}. We calculated a spin/orbital moment of $3.344/0.034\mu_B$. 

We have then used our approach to determine the exchange and anisotropy tensors of the monolayer. Figure  \ref{FigCrTe2} (b)
shows the variation of the isotropic exchange constants $J^H$ as a function of the lattice constant for the first few neighboring atomic shells.
Importantly, the figure shows how the dominant first-neighbors exchange constant switches from antiferromagnetic at the 
equilibrium lattice constant to ferromagnetic starting with a strain slightly larger than 1 \%.
We have found that the moduli of the DM vectors are negligible, of the order of a few $\mu$eV. In contrast, the intra-atomic anisotropy 
is three orders of magnitude larger and prefers an in-plane direction of the magnetization: 
$K^{zz}-K^{xx}=K^{zz}-K^{yy}=1.2$ meV. Finally, $K^{yy}-K^{xx}$ is again negligible.

\begin{figure}[ht]
\centering
\includegraphics[width=\columnwidth]{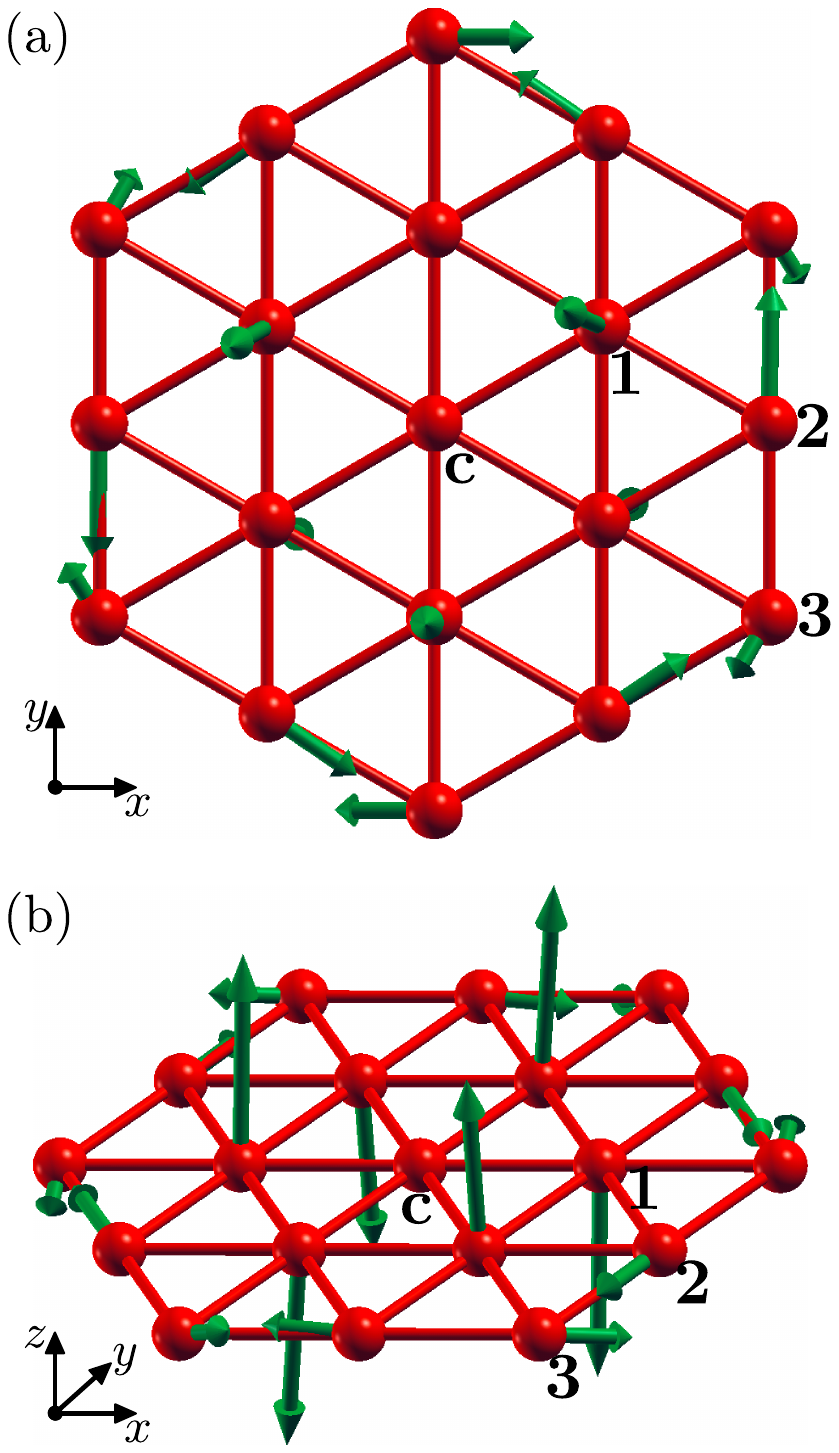}
\caption{(a) Top and (b) side views of the lower monolayer of a T-CrTe$_2$ bilayer, where only the Cr atoms 
are drawn for clarity. We show the DM vectors up to the $3^\mathrm{rd}$ neighbor. The largest moduli correspond 
to the first neighbors and are of the order of 0.2 meV. }
\label{Fig-bilayer}
\end{figure}

Finally, we have simulated a T-CrTe$_2$ bilayer with AA stacking. We have found that the interlayer  
Cr-Cr distance is 6.62 \AA, and that the magnetic moments are very similar to the monolayer case.
We have also found a somewhat reduced easy-plane intra-atomic anisotropy of $K^{zz}-K^{xx}=K^{zz}-K^{yy}=0.7$ meV. 
The intralayer Heisenberg exchange constants with respect to the atom labeled as "c" in figure 
\ref{FigCrTe2} (a) are $J_{c1}^H=1.9$ meV,  $J_{c2}^H=-4.4$ meV and $J_{c1}^H=-1.2$ meV. The symmetric intralayer exchange tensors measured in meV are
\begin{eqnarray}
\label{bilayertensors2}
 J_{c1}^{S}&=&\begin{pmatrix}
   -0.1   &-0.2  &-0.3\\
   -0.2   & 0.1    & 0.5\\
   -0.3   & 0.5  & 0.1
\end{pmatrix} \\
J_{c2}^{S}&=&\begin{pmatrix}
   -0.1 &   0     &  0.1 \\
    0   &   0   &  0   \\
    0.1 &   0     & 0
\end{pmatrix}\nonumber
\\
 J_{c3}^{S}&=&\begin{pmatrix}
   -0.1   &    0.02   &   -0.003\\
    0.02  &   -0.1    &   -0.003\\
   -0.003 &   -0.003  &   0.1
\end{pmatrix}\nonumber 
\end{eqnarray}
where the atoms 1, 2, and 3 are also labeled in the same figure. The interlayer symmetric exchange tensors 
turned out to be proportional to the identity matrix, with Heisenberg constants $J^H_{cc}=-2.2$ meV, $J^H_{c1}=-1.2$ meV and $J^H_{c2}=0.8$ meV,
meaning that the interlayer coupling is mainly ferromagnetic.
The calculated intralayer DM vectors are larger than for the monolayer compound, with moduli of about $0.2$ meV. Figure \ref{Fig-bilayer}
shows the pattern of DM vectors $D_{ci}$ where $i$ denotes the neighboring shells. We found that the interlayer DM vectors are very small,
with moduli in the $\mu$eV range again.

\section{Conclusions}
We have discussed in detail an approach to map the energy of magnetic systems within density functional theory using a 
non-orthogonal Pseudo-atomic basis set to generalized Heisenberg Hamiltonians. 
We have proposed the definition of a local operator that is suitable for both orthogonal and non-orthogonal basis sets. We have demonstrated that a correct LKAG method involves performing infinitesimal rotations only on the
exchange field, instead of full Hamiltonian. We have also shown that the rotations must be applied only to PAOs
involving localized degrees of freedom. We have established two sum rules that must be respected by the exchange 
parameters. We have also found a quantum analog of the Steiner theorem, that relates off-site to on-site 
matrix elements of the orbital angular momentum.

To test and benchmark our method, we have first investigated isolated platinum dimers and chromium trimers, where several stringent
tests could be performed. We have then analyzed the same chromium trimer deposited on a Au (111)
substrate. Here, we have found that the direction of the DM vectors sensitively changes by varying the surface-trimer distance. We have also computed
the exchange parameters of a manganese monolayer deposited epitaxially on a W(110) substrate, where the DM 
vectors follow a vortex-like pattern. We have benchmarked these last
two heterostructures against the more established KKR approach, and have found good agreement, specially for the isotropic 
exchange interactions and the anisotropies. We established that some numerical disagreements most likely occur due to the differences in the calculated electronic structure, rather than due to the differences
between the two methodologies. We have finally analyzed a 1T-CrTe$_2$ monolayer and a bilayer, where we have found easy-plane anisotropy, 
ferromagnetic interlayer coupling and have confirmed a antiferro-ferromagnetic transition controlled by the application of strain of the order 
of 1 \%, that can be achieved experimentally. 

We have analyzed the differences between our methodology and that presented in the reference article of
the PAO-based TB2J approach \cite{TB2J} by implementing the latter one in our code.  We have found that our methodology is currently 
superior in the following number of theoretical and practical aspects:\newline
(1) The LKAG approach relies on the application
of infinitesimal rotations to the angular momentum present in the Hamiltonian. We show in Section 
\ref{sec:rotating} that those rotations must be applied to the exchange field in the KS Hamiltonian. However, 
the TB2J implementation applies infinitesimal rotations to the full KS Hamiltonian. We have tested those 
two types of rotations for small atomic clusters and found that the TB2J rotation delivers non negligible energy variations when rotating the spin around the quantization axis, which should not happen  as we checked in Section \ref{IsolatedPtdimer} and it is shown in Fig. \ref{Vxcfigure}. In contrast,
our approach does not show any of these deficiencies.\newline
(2) The projection proposed in this article fulfills the sum rules. In contrast, 
the on-site projection used by the TB2J implementation fails to fulfill. We have found a failure of about 
14 \% for the case of a platinum dimer.\newline
(3) PAO basis sets span the Hilbert space in terms of localized functions whose angular parts have 
$s-$, $p-$, $d-$ and $f-$symmetry, and whose radii parts go to zero at a different cutoff radius $R_{c,i}$ for
each PAO. Those radii can be shown to be variational parameters in the PAO approaches. We have tested the impact of $R_c$ variation and energy
optimization on the magnitude of the exchange parameters. We have found it crucial to restrict the infinitesimal
rotations to the PAOs representing localized degrees of freedom, in line with the argumentation laid down above. 
Otherwise, the exchange parameters can change even by factors of four, five or orders of magnitude. Furthermore,
application of the variational principle does not lead to improved estimates of those parameters, and cannot be used as a guiding principle. The current TB2J applies infinitesimal rotations to all PAOs or Wannier functions.\newline
(4) The approach presented in this article allows us to determine all the matrix elements of the intra-atomic
anisotropy tensor. TB2J implementation can determine only the diagonal matrix elements.\newline
(5) The determination of all the matrix elements of the exchange and 
anisotropy tensors requires in the LKAG approach performing rotations around three perpendicular quantization axes. 
The TB2J approach can only perform simulations where the quantization axis is aligned in the $z$-direction
as stated in Ref. [\onlinecite{TB2J}]. 
As a consequence the TB2J method requires performing three different DFT simulations where the lattice instead of
the quantization axis must be rotated. This contrasts to our fully non-collinear approach where a single
DFT simulation need to be performed and then the exchange field can be extracted and rotated.\newline
\section*{Acknowledgments}
G. M.-C., A. G.-F. and J. F. have been funded by Ministerio de Ciencia, Innovación y Universidades, Agencia 
Estatal de Investigación, Fondo Europeo de Desarrollo Regional via the grants PGC2018-094783 and PID2022-137078NB-I00,
and by Asturias FICYT under grant AYUD/2021/51185 with the support of FEDER funds. G. M.-C. has been supported by 
Programa ``Severo Ochoa'' de Ayudas para la investigación y docencia del Principado de Asturias. He also 
acknowledges the financial support and hospitality of the Wigner Research Centre for Physics.  
This work was supported by the Ministry of Culture and Innovation and the National Research, Development and Innovation Office within the Quantum Information National Laboratory of Hungary (Grant No. 2022-2.1.1-NL-2022-00004) and projects K133827, K131938, K142179. We thank the ”Frontline” Research Excellence Programme of the NRDIO, Grant No. KKP133827. L.O. acknowledges the support of the Bolyai Research Scholarship of MTA and the Bolyai Plusz Scholarship of \'UNKP. This project has received funding from the HUN-REN Hungarian Research Network.\\


%
\end{document}